\documentclass[11pt,a4paper]{article}

\usepackage[margin=2.5cm]{geometry}
\usepackage{amsmath,amssymb,amsthm,mathtools}
\usepackage{mathrsfs}
\usepackage{enumitem}
\usepackage{graphicx}
\usepackage{xcolor}
\usepackage[colorlinks=true,linkcolor=blue,citecolor=blue,urlcolor=blue]{hyperref}

\newcommand{\R}{\mathbb R}
\newcommand{\QQ}{\mathbb Q}
\newcommand{\sQ}{\mathscr Q}
\newcommand{\sK}{\mathscr K}
\newcommand{\sM}{\mathscr M}
\newcommand{\sB}{\mathscr B}
\newcommand{\one}{\mathbf 1}
\newcommand{\intr}{\operatorname{int}}
\newcommand{\ld}{\operatorname{ld}}
\newcommand{\cd}{\operatorname{cd}}
\newcommand{\loss}{\operatorname{loss}}
\newcommand{\norm}[1]{\lVert #1\rVert}
\newcommand{\bx}{\mathbf x}
\newcommand{\by}{\mathbf y}
\newcommand{\bz}{\mathbf z}
\newcommand{\bg}{\mathbf g}

\newcommand{\bu}{\mathbf u}
\newcommand{\bp}{\mathbf p}
\newcommand{\bq}{\mathbf q}
\newcommand{\bc}{\mathbf c}
\newcommand{\be}{\mathbf e}
\newcommand{\ba}{\mathbf a}
\newcommand{\bw}{\mathbf w}
\newcommand{\bphi}{\boldsymbol\phi}
\newcommand{\beps}{\boldsymbol\varepsilon}

\newcommand{\blambda}{\boldsymbol\lambda}

\newtheorem{theorem}{Theorem}[section]
\newtheorem{proposition}[theorem]{Proposition}
\newtheorem{lemma}[theorem]{Lemma}

\theoremstyle{definition}
\newtheorem{definition}[theorem]{Definition}
\newtheorem{remark}[theorem]{Remark}
\newtheorem{example}[theorem]{Example}

\usepackage{authblk}

\makeatletter
\newcommand{\AB@affilfootnotes}{}
\DeclareRobustCommand{\AB@printaffilfootnotes}{\AB@affilfootnotes}
\g@addto@macro\@thanks{\AB@printaffilfootnotes}

\renewcommand{\affil}[2][]{%
  \g@addto@macro\AB@affilfootnotes{%
    \begingroup
      \renewcommand{\thefootnote}{#1}%
      \footnotetext[0]{#2}%
    \endgroup
  }%
}

\renewcommand{\@author}{%
  \def\rlap##1{##1}
  \AB@authlist
}
\makeatother

\title{Exact certification for conic optimization over quadratic modules:\\
partitions of unity and recession directions}
\author[1,2]{Didier Henrion}
\author[1]{Alaa Ibrahim}
\author[1,3]{Victor Magron}
\affil[1]{LAAS-CNRS, University of Toulouse, France.}
\affil[2]{Faculty of Electrical Engineering, Czech Technical University in Prague.}
\affil[3]{Institute of Mathematics of Toulouse, University of Toulouse, France.}

\date{\today}

\begin{document}
\maketitle

\begin{abstract}
We consider linear conic problems whose cone is a product of truncated quadratic modules, the sum-of-squares (SOS) side of the moment-SOS hierarchy. A floating-point solution is turned into an exact rational certificate in three steps: rounding, repair of each block residual by a constant, and absorption of the constants along a recession direction of the problem. The repair relies on strict partitions of unity in the quadratic modules. We characterize their existence, relate the canonical one to the analytic center of the dual pseudo-moment cone, Christoffel-Darboux polynomials and Pell equations, and we derive closed-form rational partitions for balls and boxes. The loss incurred by the certification is the optimal value of a recession problem, equal to a support function of the masses of the dual feasible pseudo-moments. The method is illustrated on polynomial optimization, volume computation and positivity of polynomials along D-finite trajectories.
\end{abstract}

\section{Introduction}

	Semidefinite relaxations, key components of the moment/sum-of-squares (moment/SOS)  hierarchy for polynomial
	optimization \cite{hkl20}, are usually solved in floating-point arithmetic.
	Their computed solutions satisfy the constraints only approximately, so a
	reported objective value need not be a valid bound. Certification requires
	restoring exact feasibility or rigorously controlling the effect of the
	residuals. Existing approaches can be grouped by the mechanism used to
	achieve this.
	
	\emph{Rounding, projection and exact recovery.}
	A numerical Gram matrix can be rounded to rational entries and projected
	onto the affine space defined by the coefficient identities \cite{pp08}.
	Newton refinement improves the numerical approximation before exact
	recovery; this approach also yields certificates based on SOS of rational
	functions \cite{klyz08,klyz12}. Extensions treat sums of hermitian squares
	of noncommutative polynomials \cite{ckp15}.
	Projection preserves positive semidefiniteness when the approximation
	error is sufficiently small relative to an available definiteness margin.
	For singular solutions, recovering the relevant face before rounding and
	projection can produce exact optimal semidefinite solutions over rational or
	algebraic number fields \cite{cdl24,lei25}; this technique is also used
	for noncommutative SOS bounds \cite{klm26}.
	
	\emph{Rigorous numerical validation.}
	An alternative is to prove that an exact feasible certificate exists near
	the computed one by comparing rigorous residual bounds with a
	positive-definiteness margin. This strategy is used for quadratic and
	polynomial invariants of dynamical systems \cite{rjgf12,gar19,rvs18}.
	Verified positive-definiteness tests and rounding-error bounds
	\cite{rum06,rou16} can establish validity without explicitly
	constructing large rational Gram matrices. Padding the semidefinite problem with a small
	positive-definiteness constraint supplies a margin for these tests.
	Solver tolerances alone do not provide such a guarantee \cite{rvs18}. 	The univariate setting also admits direct algorithms for exact rational
	weighted-SOS decompositions based on real or complex root isolation and
	square-free decomposition. Their bit complexity and practical performance
	are analyzed in \cite{mss19}.
	
	\emph{Residual bounds and formal verification.}
	After converting numerical SOS factors to rational data, one can bound
	the remaining polynomial error on the domain and adjust the objective
	bound accordingly. The coefficient bound over a box in
	\cite[Section~2.2]{magw15} follows this route; it certifies pointwise
	nonnegativity without requiring the residual to belong to the original
	truncated quadratic module. Formal verification supplies an additional
	layer of assurance by checking the identities and inequalities in a proof
	assistant. This is developed for polynomial and nonlinear optimization in
	\cite{mag13,mag14,magw15} and for numerical validation in
	\cite{rou16}.
	
	\emph{Dual-cone certificates.}
	Optimization directly over SOS cones and their duals can avoid a lifted
	semidefinite formulation \cite{py19}. Suitable vectors in the dual
	weighted-SOS cone certify primal membership and allow recovery of exact
	rational weighted-SOS representations \cite{dp22}. Rational algorithms
	with bounds on certificate bit size are developed in \cite{dp24}.
	The construction extends to convex cones whose duals admit suitable
	logarithmically homogeneous self-concordant barriers \cite{lpv26}.

	\emph{Perturbation and compensation.}
	Reserving an interior SOS perturbation before numerical solution makes
	it possible to absorb the residual coefficients into an exact rational
	decomposition \cite{ms18,msput}. A related a posteriori procedure rounds
	and projects a numerical Gram matrix, then lifts it into the positive
	semidefinite cone. In \cite{nwma25}, an explicit SOS decomposition of a
	constant compensates this lift and yields certified bounds in
	noncommutative polynomial optimization. The compensation cost depends
	on the spectral defect and the chosen constant decomposition.
	
	\paragraph{Contributions.}
	Building on perturbation and compensation, we develop a certification
	framework for linear conic problems over products of truncated  \emph{quadratic
	modules}. Such modules consist of semidefinite representable weighted SOS polynomials with 
	prescribed degree bounds. They are used to certify positivity on semialgebraic sets. The main advance is to connect the repair of each cone block
	with a quantified correction of the unknown through the problem's linear
	constraints. A \emph{strict partition of unity} converts each block
	residual into a constant error budget. We characterize its existence at
	the chosen relaxation order by compactness of the normalized pseudo-moment
	body, a condition stronger than the Archimedean property. Under a
	nonempty-interior assumption, the analytic center of that body and
	Christoffel--Darboux polynomials define a canonical partition.
	For balls and boxes we give explicit
	rational partitions with optimal smallest Gram eigenvalue in the
	specified normalized monomial basis; for a ball, the resulting
	constant-decomposition factor improves a dimension-based factor \cite{nwma25}.
	A \emph{recession direction} then absorbs the blockwise constants into
	the unknown, extending scalar-bound compensation to polynomial unknowns
	and coupled constraints, including differential constraints.
	Under these two structural hypotheses we prove strong duality,
	extending \cite{jh16}, and identify the optimal value of the recession
	problem as the support function of the dual feasible mass vectors.
	This characterizes the loss of the correction and guides the construction
	of explicit directions. With rational partitions and directions, the
	postprocessing yields exact rational certificates without requiring a
	strictly positive definite numerical Gram matrix. Polynomial optimization,
	volume computation and positivity along D-finite trajectories illustrate
	the construction and its certification losses.

\paragraph{Outline.}
Section~\ref{sec:conic} introduces truncated quadratic modules, pseudo-moment cones and the primal-dual pair of conic problems. Section~\ref{sec:partition} is devoted to partitions of unity: existence, analytic center, closed forms for a ball and a box, and optimality. Section~\ref{sec:repair} describes the rounding of a floating-point solution and the repair of its residuals. Section~\ref{sec:recession} introduces recession directions and proves strong duality and the formula for the optimal loss. Section~\ref{sec:examples} treats polynomial optimization, volume computation without and with Stokes constraints, and positivity along D-finite trajectories.

\section{The conic problem}\label{sec:conic}

\subsection{Quadratic modules and pseudo-moment cones}

Let $\bx\in\R^n$ and let $\bphi_r(\bx)$ be a basis with rational coefficients of $\R[\bx]_r$, the vector space of polynomials of degree up to $r$, of size $N_r:=\binom{n+r}{n}$. The cone of sums of squares (SOS) of polynomials of degree up to $r$ is
$\Sigma_r:=\{\bphi_r^TG\bphi_r : G\succeq0\}\subset\R[\bx]_{2r}$, where $G \succeq 0$ means that the symmetric Gram matrix $G$ is positive semidefinite.

Given a tuple $\bg=(g_1,\dots,g_m)$ of nonzero polynomials,
let $g_0:=1$, $d_0:=0$ and
$d_j:=\lceil\deg g_j/2\rceil$ for $j=1,\dots,m$.
Define the basic semialgebraic set
\[
\sK(\bg):=\{\bx\in\R^n:g_j(\bx)\ge0,\ j=1,\dots,m\}.
\]
For integer $r\ge\max_j d_j$, define the order-$r$
\emph{truncated quadratic module}
\[
\sQ_r(\bg):=\Big\{\sum_{j=0}^mg_js_j:\ s_j\in\Sigma_{r-d_j}\Big\}\subset\R[\bx]_{2r},
\qquad \sQ(\bg):=\bigcup_{r\ge\max_j d_j}\sQ_r(\bg).
\]
The quadratic module $\sQ(\bg)$ is \emph{Archimedean} if $R^2-\norm{\bx}^2\in\sQ(\bg)$ for some $R \in \R$. In particular, this implies that $\sK(\bg)$ is compact.

A vector $\by=(y_\ba)_{|\ba|\le2r}\in\R^{N_{2r}}$ defines the Riesz functional
\[
\ell_\by:\R[\bx]_{2r}\to\R,\qquad \ell_\by\Big(\sum_{|\ba|\le2r}p_\ba\,\bx^\ba\Big):=\sum_{|\ba|\le2r}p_\ba\,y_\ba,
\]
and every linear functional on $\R[\bx]_{2r}$ is of this form. In particular $\ell_\by(1)=y_{\mathbf 0}$ is called the mass of the pseudo-moment vector $\by$. Under this identification, the dual cone of $\sQ_r(\bg)$ is the pseudo-moment cone
\[
\sM_r(\bg):=\{\by\in\R^{N_{2r}}:\ \ell_\by(q)\ge0,\ \ \forall q\in\sQ_r(\bg)\}=\{\by\in\R^{N_{2r}}:\ M_{r-d_j}(g_j\,\by)\succeq0,\ j=0,\dots,m\},
\]
where $M_{r-d_j}(g_j\,\by):=\ell_\by(g_j\,\bphi_{r-d_j}\bphi_{r-d_j}^T)$, the Riesz functional being applied entrywise, is the moment ($j=0$) or localizing ($j\ge1$) matrix. The second description follows from the identity
\begin{equation}\label{eq:pairing}
\ell_\by(g_j\,\bphi_{r-d_j}^TG\,\bphi_{r-d_j})=\langle G,M_{r-d_j}(g_j\,\by)\rangle 
\end{equation}
where $\langle A,B \rangle := \mathrm{trace}(A^T B)$ denotes the inner product between two matrices.

For a convex set $C$ in a finite-dimensional real vector space $E$,
$\intr C$ denotes its interior in $E$, whereas its \emph{relative interior}
$\operatorname{ri}C$ is its interior in the affine hull of $C$.
Thus $\operatorname{ri}C=\intr C$ when $C$ is full-dimensional.
For example, in the space of symmetric matrices,
$\operatorname{ri}\{G\succeq0\}=\{G\succ0\}$.
We use the identities
$\operatorname{ri}(A(C))=A(\operatorname{ri}C)$, 
$\operatorname{ri}(C+D)=\operatorname{ri}C+\operatorname{ri}D$,
for a linear map $A$ and nonempty convex sets $C,D$;
see \cite[Theorem~6.6 and Corollary~6.6.2]{roc70}.
Interiors of polynomial cones below are taken in their ambient polynomial spaces.

\begin{lemma}\label{lem:int}
A polynomial $p$ belongs to $\intr\Sigma_r$ if and only if
$p=\bphi_r^TG\bphi_r$ for some $G\succ0$.
If moreover $p\in\QQ[\bx]$, then $G$ can be chosen rational.
\end{lemma}

\begin{proof}
This is the Gram-map instance of \cite[Theorem~6.6]{roc70}.
The Gram map $G\mapsto\bphi_r^TG\bphi_r$ is linear and onto $\R[\bx]_{2r}$, hence open, so it maps the open cone $\{G\succ0\}$ into $\intr\Sigma_r$. Conversely, if $p\in\intr\Sigma_r$ then $p-\varepsilon\bphi_r^T\bphi_r\in\Sigma_r$ for some $\varepsilon>0$, so $p$ has a Gram matrix $\varepsilon I+G'\succ0$. 
{For the second claim, see \cite[Theorem~1.2]{hil09}.}
\end{proof}

\subsection{Primal-dual pair}

We consider linear conic problems with $m$ blocks. Block $k$ has its own variables $\bx\in\R^{n_k}$ and a tuple $\bg_k=(g_{k,1},\dots,g_{k,m_k})$ of generators with rational coefficients, with $g_{k,0}:=1$, $d_{k,j}:=\lceil\deg g_{k,j}/2\rceil$. Let
\[
V_r:=\prod_{k=1}^m\R[\bx]_{2r},\qquad \sQ_r:=\prod_{k=1}^m\sQ_r(\bg_k)\subset V_r,\qquad \sM_r:=\prod_{k=1}^m\sM_r(\bg_k).
\]
For $\by=(\by_1,\dots,\by_m)$ and $\be=(e_1,\dots,e_m)\in V_r$ let $\ell_\by(\be):=\sum_k\ell_{\by_k}(e_k)$. Every linear functional on $V_r$ is of this form, and $\sM_r$ is the dual cone of $\sQ_r$.

The unknown $\bp$ lives in a real vector space $P$ of tuples of polynomials. The data are a linear map $A:P\to\prod_k\R[\bx]$, a linear functional $b:P\to\R$ and $\bc=(c_1,\dots,c_m)\in\prod_k\QQ[\bx]$. Both $A$ and $b$ may involve differentiation, integration against measures with rational moments, or point evaluation at rational points. Let $P_r:=\{\bp\in P:\deg\bp\le2r,\ A(\bp)\in V_r\}$, a finite-dimensional space, and assume that $\bc\in V_r$ and that $A$ and $b$ have rational matrices in a rational basis of $P_r$. The order-$r$ problem is
\begin{equation}\label{eq:primal}
v_r:=\sup_{\bp\in P_r}\ b(\bp)\quad\text{s.t.}\quad \bc-A(\bp)\in\sQ_r,
\end{equation}
with dual
\begin{equation}\label{eq:dual}
\min_{\by\in\sM_r}\ c(\by)\quad\text{s.t.}\quad A^*(\by)=b,
\end{equation}
where $c(\by):=\ell_\by(\bc)$ and the adjoint map $A^*(\by)\in P_r^*$ is defined by $A^*(\by)(\bp):=\ell_\by(A(\bp))$. Weak duality is immediate: if $\bp$ and $\by$ are feasible then
\begin{equation}\label{eq:weak}
b(\bp)=\ell_\by(A(\bp))=c(\by)-\ell_\by(\bc-A(\bp))\le c(\by).
\end{equation}
Under some mild assumptions (to be covered) there is no duality gap and the minimum in \eqref{eq:dual} is attained, so $v_r$ denotes the common value of both primal and dual problems.

Problem \eqref{eq:primal} is an inner approximation. Since elements of $\sQ(\bg_k)$ are nonnegative on $\sK(\bg_k)$, the computation \eqref{eq:weak} with $\by$ replaced by a tuple of measures shows that
\[
v_r\le v_{r+1}\le v^\star:=\inf\Big\{\sum_k\int c_k\,d\mu_k:\ \sum_k\int A_k(\bp)\,d\mu_k=b(\bp),\ \ \forall\bp\in P\Big\},
\]
the infimum being over tuples of positive measures $\mu_k$ supported on $\sK(\bg_k)$.
Consequently, \emph{every exactly feasible $\bp$ certifies the lower bound $b(\bp)\le v_r\le v^\star$}. The purpose of this work is to produce such a $\bp$, with rational certificates, from a floating-point solution of \eqref{eq:primal}.

\section{Partitions of unity}\label{sec:partition}

In this section $\bg=(g_1,\dots,g_m)$ is a tuple of polynomials in $\bx\in\R^n$ with $\sK(\bg)\neq\emptyset$. Recall that $g_0 \equiv 1$.

\subsection{Definition and existence}

\begin{definition}
A \emph{partition of unity of order $r$} in $\sQ_r(\bg)$ is a tuple $\bu=(u_0,\dots,u_m)$ with $u_j\in\Sigma_{r-d_j}$ and
\begin{equation}\label{eq:pou}
1=\sum_{j=0}^mg_ju_j.
\end{equation}
It is \emph{strict} if $u_0\in\intr\Sigma_r$.
\end{definition}

Let $\sB_r(\bg):=\{\by\in\sM_r(\bg):\ell_\by(1)=1\}$ denote the \emph{pseudo-moment body}, the set of normalized pseudo-moments. It contains the moment vectors of the Dirac measures at points of $\sK(\bg)$.

\begin{proposition}\label{prop:H}
The following are equivalent:
\begin{enumerate}[label=(\alph*),topsep=0pt,itemsep=0pt]
\item there is a strict partition of unity of order $r$ in $\sQ_r(\bg)$;
\item $1\in\intr\sQ_r(\bg)$;
\item if $\by\in\sM_r(\bg)$ and $\ell_\by(1)=0$ then $\by=\mathbf 0$;
\item $\sB_r(\bg)$ is compact.
\end{enumerate}
If they hold for a sub-tuple of $\bg$, they hold for $\bg$.
\end{proposition}

\begin{proof}
{
Write $Q:=\sQ_r(\bg)$, $M:=\sM_r(\bg)$ and
$B:=\sB_r(\bg)$. The cone $Q$ is full-dimensional because it contains
$\Sigma_r$, which has nonempty interior by Lemma~\ref{lem:int}.
The dual cone $M$ is closed, and $\ell_\by(1)\ge0$ for every $\by\in M$.
Also $B\neq\emptyset$, since $\sK(\bg)\neq\emptyset$.
 
\emph{(a)$\Leftrightarrow$(b).}
Let $\mathscr R:=\{\sum_{j=1}^m g_js_j:s_j\in\Sigma_{r-d_j}\}$.
Then $Q=\Sigma_r+\mathscr R$, and the relative-interior sum rule gives
\[
\intr Q=\intr\Sigma_r+\operatorname{ri}\mathscr R
\subseteq\intr\Sigma_r+\mathscr R.
\]
Conversely, $\intr\Sigma_r+\mathscr R$ is an open subset of $Q$,
so it is contained in $\intr Q$. Thus
$1\in\intr Q$ is equivalent to a decomposition
$1=u_0+\sum_{j=1}^m g_ju_j$ with
$u_0\in\intr\Sigma_r$ and $u_j\in\Sigma_{r-d_j}$.
 
\emph{(b)$\Rightarrow$(c).}
For every $q\in\R[\bx]_{2r}$ there is $\varepsilon>0$ such that
$1\pm\varepsilon q\in Q$. If $\by\in M$ has zero mass, applying
$\ell_\by$ to both polynomials yields $\pm\varepsilon\ell_\by(q)\ge0$.
Hence $\ell_\by(q)=0$ for every $q$, and $\by=\mathbf0$.
 
\emph{(c)$\Rightarrow$(b).}
Suppose $1\notin\intr Q$. Separating $1$ from the nonempty open
convex set $\intr Q$ gives a nonzero functional $\ell_\by$ with
$\ell_\by(q)\ge\ell_\by(1)$ for $q\in\intr Q$
\cite[Theorem~11.3]{roc70}.
Since $Q\subseteq\operatorname{cl}(\intr Q)$, this inequality holds on $Q$.
Taking $q=0$ and $q=2$ shows $\ell_\by(1)=0$; the same inequality
then gives $\by\in M$, contradicting (c).
 
\emph{(c)$\Rightarrow$(d).}
The set $B=M\cap\{\by:\ell_\by(1)=1\}$ is closed.
If it were unbounded, there would be $\by^{(i)}\in B$ with
$\norm{\by^{(i)}}\to\infty$. After passage to a subsequence,
$\bz^{(i)}:=\by^{(i)}/\norm{\by^{(i)}}$ converges to a vector
$\bz\in M$ with $\norm{\bz}=1$ and
\[
\ell_\bz(1)=\lim_i\frac{1}{\norm{\by^{(i)}}}=0,
\]
contradicting (c). Thus $B$ is closed and bounded, hence compact.
 
\emph{(d)$\Rightarrow$(c).}
Choose $\bz \in B$. If a nonzero $\by\in M$ had zero mass, then
$\bz+t\by\in B$ for every $t\ge0$, contradicting the boundedness of $B$.
 
Finally, adding generators enlarges $Q$ in the same ambient space,
so property (b), and therefore all four properties, is preserved.
}
\end{proof}

\begin{remark}[Archimedean is not enough]\label{rem:arch}
Describe the box $\prod_i[\alpha_i-\rho_i,\alpha_i+\rho_i]$ by the $2n$ linear generators $1\pm z_i$, where $z_i:=(x_i-\alpha_i)/\rho_i$. The identity
$1-z_i^2=\tfrac12\big((1+z_i)^2(1-z_i)+(1-z_i)^2(1+z_i)\big)$
shows that $n-\norm{\bz}^2\in\sQ_2(\bg)$, so $\sQ(\bg)$ is Archimedean, this property being invariant under the affine change of coordinates $\bx\mapsto\bz$. Yet no strict partition of unity exists at any order $r\ge1$: each product $g_ju_j$, $j\ge1$, has degree at most $2r-1$, so \eqref{eq:pou} forces the degree-$2r$ form of $u_0$ to vanish, whereas by Lemma~\ref{lem:int} this form is positive definite when $u_0\in\intr\Sigma_r$. The same holds for a simplex with linear generators. The relevant hypothesis is therefore not the Archimedean property but its quantitative, order-$r$ version (b). It holds at every order when $\bg$ contains the quadratic generators of a ball or of a box, see Section~\ref{sec:closed}.
\end{remark}

\subsection{Analytic center and the dual pseudo-moment cone}

Let $N:=\sum_{j=0}^mN_{r-d_j}$ be the total size of the Gram matrices in \eqref{eq:pou}. For $\by\in\R^{N_{2r}}$ with positive definite moment and localizing matrices, define
\[
\ld_r(\by):=\sum_{j=0}^m\log\det M_{r-d_j}(g_j\,\by) \in \R,\qquad
\cd_r(\by):=\sum_{j=0}^mg_j\,\bphi_{r-d_j}^TM_{r-d_j}(g_j\,\by)^{-1}\bphi_{r-d_j}\ \in\R[\bx]_{2r}.
\]
Each term of $\cd_r(\by)$ is $g_j$ times the Christoffel--Darboux polynomial (reciprocal of the Christoffel function) of the functional $q\mapsto\ell_\by(g_jq)$, see \cite[Section~4.1]{lpp22}.

\begin{theorem}[Analytic center]\label{thm:ac}
Assume that the conditions of Proposition~\ref{prop:H} hold and that $\sK(\bg)$ has nonempty interior. Then:
\begin{enumerate}[label=(\roman*),topsep=0pt,itemsep=0pt]
\item $\ld_r$ has a unique maximizer $\by^\star$ on $\sB_r(\bg)$, the \emph{analytic center} of $\sB_r(\bg)$;
\item $\by^\star$ is the unique $\by\in\sB_r(\bg)$ with positive definite moment and localizing matrices such that $\cd_r(\by)$ is a constant polynomial, and then $\cd_r(\by^\star)\equiv N$;
\item the polynomials
\begin{equation}\label{eq:acpou}
u_j^\star:=\tfrac1N\,\bphi_{r-d_j}^TM_{r-d_j}(g_j\,\by^\star)^{-1}\bphi_{r-d_j},\qquad j=0,\dots,m,
\end{equation}
form a strict partition of unity, and their Gram matrices are the unique maximizer of $\sum_j\log\det G_j$ among the Gram matrices $(G_j)_j$ of partitions of unity of order $r$.
\end{enumerate}
\end{theorem}

\begin{proof}
(i) The conditions of Proposition~\ref{prop:H} imply that
$\sK(\bg)$ is compact: the moment vectors of its Dirac measures
belong to the bounded set $\sB_r(\bg)$, which bounds the
coordinates $\ell_\by(x_i^2)$.
Since $\sK(\bg)$ has nonempty interior, its normalized Lebesgue
measure is well defined.

For every $j$ and every nonzero polynomial
$q\in\R[\bx]_{r-d_j}$, the polynomial $g_jq^2$ is nonnegative
on $\sK(\bg)$ and positive almost everywhere in its interior.
Indeed, neither $g_j$ nor $q$ is identically zero, and the
zero set of a nonzero polynomial has Lebesgue measure zero.
Consequently, $\int_{\sK(\bg)}g_jq^2\,d\bx>0$.
Thus there exists $\bar\by\in\sB_r(\bg)$ for which all moment
and localizing matrices are positive definite.

Extend $\ld_r$ by $-\infty$ whenever one of these matrices
is singular. On the compact set $\sB_r(\bg)$, all matrices
are bounded, and $\ld_r$ tends to $-\infty$ whenever a
sequence of positive definite tuples approaches a tuple
with a singular matrix. Hence $\ld_r$ attains its maximum
at a point with positive definite matrices.

The map $\by\mapsto M_r(\by)$ is injective because products
of polynomials of degree at most $r$ span
$\R[\bx]_{2r}$. Strict concavity of $\log\det$ (see, e.g., \cite[§3.1.5, pp. 74–75]{boyd04}) therefore
implies uniqueness.

{
(ii) Write $L_j(\by):=M_{r-d_j}(g_j\,\by)$.
For any direction $\by'$, linearity gives
$L_j(\by+t\by')=L_j(\by)+tL_j(\by')$.
When all $L_j(\by)\succ0$, these matrices remain positive definite for
sufficiently small $|t|$. The differential identity
$\frac{d}{dt}\log\det(X+tH)|_{t=0}=\operatorname{trace}(X^{-1}H)$
therefore yields
\begin{align*}
D\ld_r(\by)[\by']
&=\left.\frac{d}{dt}\right|_{t=0}\sum_{j=0}^m\log\det L_j(\by+t\by')\\
&=\sum_{j=0}^m\langle L_j(\by)^{-1},L_j(\by')\rangle
=\ell_{\by'}(\cd_r(\by)),
\end{align*}
where the last equality follows by applying \eqref{eq:pairing} to each summand.
 
The affine constraint defining the normalized slice is $\ell_\by(1)=1$,
so its tangent directions are precisely the $\by'$ with $\ell_{\by'}(1)=0$.
At a positive definite point both signs of every sufficiently small tangent
perturbation are feasible. A maximizer must therefore satisfy
\[
\ell_{\by'}(\cd_r(\by))=0\qquad\text{whenever }\ell_{\by'}(1)=0.
\]
Conversely, the above stationarity condition is sufficient by concavity: for every other positive
definite $\bz\in\sB_r(\bg)$,
\[
\ld_r(\bz)\le\ld_r(\by)+D\ld_r(\by)[\bz-\by]=\ld_r(\by),
\]
and points with a singular matrix have value $-\infty$.
\if{
The annihilator of the hyperplane $\{\by':\ell_{\by'}(1)=0\}$ is the set of polynomials $p$ such that $\ell_{\by'}(p) = 0$ for all $\by'$ with $\ell_{\by'}(1)=0$. 
Thus, the annihilator is the one-dimensional space spanned by the constant polynomial $1$.
Since the coefficient--moment pairing is nondegenerate, the stationarity
condition is equivalent to $\cd_r(\by)\equiv\lambda$ for some scalar $\lambda$.
}\fi
For each \(\alpha\neq\mathbf0\), choose \(\mathbf y'\) with \(y'_\alpha=1\) and all other coordinates zero. Then \(\ell_{\mathbf y'}(1)=0\), and stationarity forces the coefficient of \(\mathbf x^\alpha\) in \(\operatorname{cd}_r(\mathbf y)\) to vanish.
Thus all nonconstant coefficients vanish and $\cd_r(\by)\equiv\lambda$ for some scalar $\lambda$. 
Finally,
\[
\lambda=\ell_\by(\cd_r(\by))
=\sum_{j=0}^m\operatorname{trace}\big(L_j(\by)^{-1}L_j(\by)\big)
=\sum_{j=0}^m N_{r-d_j}=N,
\]
where the first equality uses $\ell_\by(1)=1$.
Uniqueness follows from (i).
}

(iii) By (ii), $\sum_jg_ju^\star_j=1$ with Gram matrices $G_j^\star:=M_{r-d_j}(g_j\,\by^\star)^{-1}/N\succ0$. If $(G_j)_j$ are Gram matrices of a partition of unity, then $\sum_j\langle G_j,M_{r-d_j}(g_j\,\by^\star)\rangle=\ell_{\by^\star}(1)=1$ by \eqref{eq:pairing}, and concavity of $\log\det$ yields
\[
\sum_j\log\det G_j\le\sum_j\log\det G^\star_j+N\sum_j\langle M_{r-d_j}(g_j\,\by^\star),G_j-G_j^\star\rangle=\sum_j\log\det G^\star_j .
\]
Uniqueness follows from strict concavity.
\end{proof}

This primal-dual log-det pair goes back to \cite{nes00} and appears in \cite{dghl21}, \cite{las23} and \cite[Theorem~3.2]{lx24}; the proof above is self-contained.

\begin{example}[Interval]\label{ex:interval}
Let $n=1$ and $g_1=1-x^2$. Let $T_i$ and $U_i$ be the Chebyshev polynomials of the first and second kind. The orthonormal polynomials of the arcsine measure $dx/(\pi\sqrt{1-x^2})$ are $1$ and $\sqrt2\,T_i$, $i\ge1$, and those of its product with $g_1$ are $\sqrt2\,U_i$, $i\ge0$. The Chebyshev polynomials satisfy the polynomial Pell equation
\[
T_i(x)^2+(1-x^2)U_{i-1}(x)^2=1,\qquad i\ge1,
\]
see, e.g., \cite[\S 1.1]{las23}.
If $\by$ is the vector of arcsine moments, summing the above equations gives
\[
\cd_r(\by)=1+2\sum_{i=1}^r\big(T_i^2+(1-x^2)U_{i-1}^2\big)\equiv 2r+1 \equiv N .
\]
By Theorem~\ref{thm:ac}(ii), the analytic center is the vector of arcsine moments at every order, and \eqref{eq:acpou} is the rational partition
\[
1=\frac1{2r+1}\Big(1+2\sum_{i=1}^r T_i(x)^2\Big)+(1-x^2)\,\frac2{2r+1}\sum_{i=0}^{r-1}U_i(x)^2 .
\]
\end{example}

\begin{remark}
For the unit ball with the single generator $1-\norm{\bx}^2$, the generalized Pell equation \cite[Theorem~4.1]{lx24} states that $\cd_r(\by)$ is constant when $\by$ is the moment vector of the equilibrium measure, proportional to $d\bx/\sqrt{1-\norm{\bx}^2}$. By Theorem~\ref{thm:ac}(ii), the analytic center is then this moment vector at every order. For the box with the $n$ generators $1-x_i^2$ the analytic center exists, but it is not the moment vector of the equilibrium measure; the Pell equation requires products of generators \cite[Theorem~4.5]{lx24}. In general $\by^\star$ is computed numerically and is not rational. For certification, any strict rational partition will do, and we now derive closed-form ones.
\end{remark}

\subsection{Closed-form partitions for a ball and a box}\label{sec:closed}

Let $\alpha_i\in\QQ^n$ and $\rho_i\in\QQ_{>0}$,
and let $\bz=(z_1,\dots,z_n)$ with $z_i:=(x_i-\alpha_i)/\rho_i$ be normalized coordinates. We take $\bphi_r:=(\bz^\ba)_{|\ba|\le r}$ as a basis for $\R[\bx]_r$ since $\bx\mapsto\bz$ is affine and invertible.

\begin{theorem}[Ball]\label{thm:ball}
Let $\rho_i=\rho$ with $\rho>0\in\QQ$, and $g:=1-\norm{\bz}^2$. For every $r\ge1$,
\[
u_0:=\frac1{r+1}\sum_{i=0}^r\norm{\bz}^{2i},\qquad u_1:=\frac1{r+1}\sum_{i=0}^{r-1}(r-i)\norm{\bz}^{2i}
\]
form a strict rational partition of unity $1=u_0+g\,u_1$ of order $r$. In the basis $\bphi_r$, the Gram matrix of $u_0$ is diagonal with entries $|\ba|!/(\ba!\,(r+1))\ge1/(r+1)$.
\end{theorem}

\begin{proof}
Summing $1-\norm{\bz}^{2i}=(1-\norm{\bz}^2)\sum_{l=0}^{i-1}\norm{\bz}^{2l}$ over $i=0,\dots,r$ gives
$(r+1)-\sum_{i=0}^r\norm{\bz}^{2i}=(1-\norm{\bz}^2)\sum_{i=0}^{r-1}(r-i)\norm{\bz}^{2i}$.
By the multinomial formula $\norm{\bz}^{2i}=\sum_{|\ba|=i}\frac{i!}{\ba!}\bz^{2\ba}$, so $u_0$ and $u_1$ are positive combinations of squares of the $\bz^\ba$, with all $\bz^\ba$, $|\ba|\le r$, appearing in $u_0$. Lemma~\ref{lem:int} applies.
\end{proof}

\begin{theorem}[Box]\label{thm:box}
Let $\rho_i>0\in\QQ$ and $\bg=(g_1,\dots,g_n)$ with $g_i:=1-z_i^2$. For every $r\ge1$,
\[
u_0:=\frac1{N_r}\sum_{|\ba|\le r}\bz^{2\ba},\qquad
u_i:=\frac1{N_r}\sum_{\substack{|\ba|\le r\\ a_i\ge1}}\Big(\prod_{l<i}z_l^{2a_l}\Big)\sum_{l=0}^{a_i-1}z_i^{2l},\quad i=1,\dots,n,
\]
form a strict rational partition of unity $1=u_0+\sum_ig_iu_i$ of order $r$. In the basis $\bphi_r$, the Gram matrix of $u_0$ is $I/N_r$.
\end{theorem}

\begin{proof}
For $|\ba|\le r$, telescoping over the coordinates and then within each coordinate gives
\[
1-\bz^{2\ba}=\sum_{i=1}^n\Big(\prod_{l<i}z_l^{2a_l}\Big)\big(1-z_i^{2a_i}\big)
=\sum_{i=1}^ng_i\Big(\prod_{l<i}z_l^{2a_l}\Big)\sum_{l=0}^{a_i-1}z_i^{2l}.
\]
Sum over $|\ba|\le r$ and divide by $N_r$. Each term of $u_i$ is the square of a monomial of degree at most $r-1$.
\end{proof}

\begin{proposition}[Optimal margins]\label{prop:optimal}
Let $1=u_0+\sum_jg_ju_j$ be a partition of unity of order $r$ for the ball of Theorem~\ref{thm:ball} or the box of Theorem~\ref{thm:box}, and let $U_0$ be a Gram matrix of $u_0$ in the basis $\bphi_r=(\bz^\ba)_{|\ba|\le r}$. Then $\lambda_{\min}(U_0)\le\frac1{r+1}$ for the ball and $\lambda_{\min}(U_0)\le\frac1{N_r}$ for the box. Hence the partitions of Theorems~\ref{thm:ball} and~\ref{thm:box} maximize $\lambda_{\min}(U_0)$.
\end{proposition}
 
\begin{proof}
At a point where all the $g_j$ vanish, $1=u_0=\bphi_r^TU_0\bphi_r\ge\lambda_{\min}(U_0)\norm{\bphi_r}^2$. For the box, take a vertex $\bz\in\{-1,1\}^n$: every monomial equals $\pm1$, so $\norm{\bphi_r(\bz)}^2=N_r$. For the ball, take the first unit vector: the nonzero entries of $\bphi_r$ are those of $z_1^i$, $i=0,\dots,r$, so $\norm{\bphi_r}^2=r+1$.
\end{proof}

The optimality statement concerns the specified normalized
monomial basis. The smallest eigenvalue of a Gram matrix
is not invariant under a general change of polynomial basis.

\begin{remark}
The constant decomposition of \cite[Lemma~5]{nwma25} has the same constant, the size of the Gram matrix, for the four constraint families considered there. By Proposition~\ref{prop:optimal} this is optimal for the box, whereas for the ball Theorem~\ref{thm:ball} reduces it from $N_r$ to $r+1$. The reduction carries over to the free algebra in self-adjoint variables $X_1,\dots,X_n$. Let $S:=\sum_iX_i^2$, let $v_r$ be the vector of all words of length at most $r$, and let $T_k:=\sum_{|w|=k}w^*w$. Since $T_{k+1}=\sum_{|w|=k}w^*Sw$, we have $1-T_{k+1}=(1-T_k)+\sum_{|w|=k}w^*(1-S)w$, and summing over $k$ gives
\[
r+1=v_r^*v_r+\sum_{j=0}^{r-1}(r-j)\sum_{|w|=j}w^*(1-S)\,w ,
\]
to be compared with the constant $(n^{r+1}-1)/(n-1)$ of \cite[Lemma~5]{nwma25}. Theorem~\ref{thm:ball} is the commutative image of this identity.
\end{remark}

\section{Rounding and repair}\label{sec:repair}

We return to \eqref{eq:primal} and assume:

\textbf{(H)} for every $k$, $1\in\intr\sQ_r(\bg_k)$.

By Proposition~\ref{prop:H} and Section~\ref{sec:closed}, \textbf{(H)} holds as soon as each $\bg_k$ contains, up to positive rational factors, the generators of a ball or of a box, and then rational strict partitions of unity
\begin{equation}\label{eq:pouk}
1=\sum_{j=0}^{m_k}g_{k,j}u_{k,j},\qquad u_{k,0}=\bphi_r^TU_k\bphi_r,\quad U_k\succ0,
\end{equation}
are available in closed form, with $u_{k,j}:=0$ for the generators not involved.

{
In block $k$, write $\bphi^{(k)}_s$ for the chosen rational basis of
polynomials in that block's $n_k$ variables of degree at most $s$.
A floating-point solver applied to \eqref{eq:primal} returns $\tilde\bp$
and Gram matrices $\tilde G_{k,j}$ such that
\[
c_k-A_k(\tilde\bp)\approx
\sum_{j=0}^{m_k}g_{k,j}\,
(\bphi^{(k)}_{r-d_{k,j}})^T\tilde G_{k,j}\bphi^{(k)}_{r-d_{k,j}},
\qquad d_{k,j}=\left\lceil\frac{\deg g_{k,j}}{2}\right\rceil.
\]
Thus $\tilde G_{k,j}$ has size
$\binom{n_k+r-d_{k,j}}{n_k}$, and the unweighted block $j=0$ uses degree $r$.
 
\textbf{Rounding.} Round the coordinates of $\tilde\bp$ in a rational basis of
$P_r$ to obtain $\hat\bp$. For every $k$ and $j=0,\dots,m_k$, factor
$\tilde G_{k,j}\approx\tilde L_{k,j}\tilde L_{k,j}^T$, round
$\tilde L_{k,j}$ to a rational matrix $\hat L_{k,j}$, and set
\[
\hat G_{k,j}:=\hat L_{k,j}\hat L_{k,j}^T\succeq0,\qquad
\hat s_{k,j}:=(\bphi^{(k)}_{r-d_{k,j}})^T\hat G_{k,j}\bphi^{(k)}_{r-d_{k,j}}.
\]
Then $\hat s_{k,j}\in\Sigma_{r-d_{k,j}}$ is an exact rational SOS.
Below we again suppress the superscript $(k)$ when the block is clear.
}

\textbf{Residuals.} For each block define the rational residual
\begin{equation}\label{eq:res}
\hat e_k:=c_k-A_k(\hat\bp)-\sum_{j=1}^{m_k}g_{k,j}\hat s_{k,j}\ \in\QQ[\bx]_{2r}.
\end{equation}
This is an exact identity in which all numerical and rounding errors are concentrated in $\hat e_k\approx\hat s_{k,0}$. In general $\hat e_k$ is not SOS: by complementarity the optimal Gram matrix of $s_{k,0}$ is singular, so $\hat e_k$ lies near the boundary of $\Sigma_r$. Let $D_k$ be a rational symmetric matrix with $\bphi_r^TD_k\bphi_r=\hat e_k-\hat s_{k,0}$, for instance the one obtained by splitting each coefficient equally among the corresponding entries.

\begin{lemma}[Repair]\label{lem:repair}
If $\varepsilon_k\in\QQ_+$ is such that $D_k+\varepsilon_kU_k\succeq0$, then $\hat e_k+\varepsilon_ku_{k,0}\in\Sigma_r$ with rational Gram matrix $\hat G_{k,0}+D_k+\varepsilon_kU_k\succeq0$. When $U_k$ is diagonal with entries at least $\delta_k$, a sufficient condition, checkable in rational arithmetic, is $\varepsilon_k^2\delta_k^2\ge\norm{D_k}_F^2$.
\end{lemma}

\begin{proof}
The Gram matrix is the sum of $\hat G_{k,0}\succeq0$ and $D_k+\varepsilon_kU_k\succeq0$. For the second claim, $\varepsilon_kU_k\succeq\varepsilon_k\delta_kI\succeq\norm{D_k}_FI\succeq-D_k$.
\end{proof}

With the partitions of Section~\ref{sec:closed}, $\delta_k=1/(r+1)$ for a ball and $\delta_k=1/N_r$ for a box. For fixed bases and fixed relaxation order, the repair
depends linearly on the mismatch:
one may choose a rational $\varepsilon_k$ arbitrarily
close from above to $\norm{D_k}_F/\delta_k$.
Thus the sufficient repair amplifies the mismatch by
$r+1$ for the ball and by $N_r$ for the box.
These factors are not uniform in the relaxation order.

\begin{remark}[Round, project, lift]\label{rem:rpl}
In the terminology of \cite{nwma25}, $\hat G_{k,0}+D_k$ is the Gram matrix output by the the rounding and projection algorithm from \cite{pp08}. 
The equal splitting is the Frobenius-optimal projection in a monomial basis \cite[Lemma~2 and Corollary~3]{nwma25}. 
Lemma~\ref{lem:repair} with $U_k=\delta_kI$ is the lifting step \cite[Theorem~4]{nwma25}, whose loss $-\mu_{\min}N_r$ corresponds to $\varepsilon_k=-\mu_{\min}/\delta_k$ with $\delta_k=1/N_r$, where $\mu_{\min}$ is a lower bound on the spectrum of the projected matrix. There are two differences. First, $\mu_{\min}$ must be certified, by interval arithmetic in \cite{nwma25}; here the factors are rounded, so that $\hat G_{k,0}\succeq0$ exactly, and the test $\varepsilon_k\delta_k\ge\norm{D_k}_F$ involves rational arithmetic only. Writing $H_k:=\hat G_{k,0}+D_k$, we have
\[
\max\{0,-\lambda_{\min}(H_k)\}\le\norm{D_k}_F,
\]
because $\hat G_{k,0}\succeq0$.
Thus the Frobenius test is conservative relative to the
exact spectral defect. A certified lower bound
$\underline\lambda_k\le\lambda_{\min}(H_k)$ gives the
alternative sufficient repair
\[
\varepsilon_k\delta_k
\ge\max\{0,-\underline\lambda_k\}.
\]
Whether this improves the Frobenius test depends on the
sharpness of the certified lower bound. Second, in \cite{nwma25} the unknown is the bound, so the constant is absorbed at once, as in Section~\ref{sec:pop} where $\loss_r(\varepsilon)=\varepsilon$; the recession problem \eqref{eq:rec} from Section \ref{sec:recession} replaces this step when the unknown is a polynomial. Conversely, \cite{nwma25} covers noncommutative variables, equality constraints through normal forms, correlative sparsity, where the losses of the cliques add up, and symmetry reduction, and it tightens the bound when the projected matrix is positive definite \cite[Theorem~6]{nwma25}, none of which is addressed here.
\end{remark}

\begin{proposition}[Repaired certificate]\label{prop:repaired}
With $\beps:=(\varepsilon_1,\dots,\varepsilon_m)$ as in Lemma~\ref{lem:repair}, identified with a tuple of constant polynomials,
\[
c_k-A_k(\hat\bp)+\varepsilon_k=(\hat e_k+\varepsilon_ku_{k,0})+\sum_{j=1}^{m_k}g_{k,j}\,(\hat s_{k,j}+\varepsilon_ku_{k,j})\quad\text{for all }k,
\]
hence $\bc+\beps-A(\hat\bp)\in\sQ_r$, with rational SOS multipliers.
\end{proposition}

\begin{proof}
Add $\varepsilon_k$ times \eqref{eq:pouk} to \eqref{eq:res}.
\end{proof}

Thus $\hat\bp$ is exactly feasible for the problem with $\bc$ relaxed to $\bc+\beps$. The multiplier of $g_{k,0}=1$ has been shifted by the interior direction $\varepsilon_ku_{k,0}$, at the price of the constant $\varepsilon_k$.

\section{Recession directions and strong duality}\label{sec:recession}

It remains to absorb $\beps$. For $\beps\in\R^m_+$ consider the \emph{recession problem}
\begin{equation}\label{eq:rec}
\loss_r(\beps):=\inf_{\bq\in P_r}\ -b(\bq)\quad\text{s.t.}\quad -A(\bq)-\beps\in\sQ_r .
\end{equation}
If $\bq$ is feasible for \eqref{eq:rec}, Proposition~\ref{prop:repaired} gives
$\bc-A(\hat\bp+\bq)=\big(\bc+\beps-A(\hat\bp)\big)+\big(-A(\bq)-\beps\big)\in\sQ_r$,
so $\hat\bp+\bq$ is exactly feasible for \eqref{eq:primal}, with loss $-b(\bq)$. We assume:

\textbf{(R)} there is $\bq^\circ\in P_r$ such that $-A(\bq^\circ)-\one\in\sQ_r$, where $\one:=(1,\dots,1)\in V_r$.

Hypothesis \textbf{(R)} guarantees the existence of a recession direction, see Appendix \ref{app:recession} for more details.

\begin{theorem}[Dual attainment]\label{thm:duality}
Assume \textbf{(H)} and \textbf{(R)}. Then:
\begin{enumerate}[label=(\roman*),topsep=0pt,itemsep=0pt]
\item problem \eqref{eq:primal} is strictly feasible: $\bc-A(\bp)\in\intr\sQ_r$ for some $\bp\in P_r$;
\item the feasible set of \eqref{eq:dual} is compact, and its elements satisfy $\sum_k\ell_{\by_k}(1)\le-b(\bq^\circ)$;
\item if \eqref{eq:dual} is feasible, then $v_r$ is finite, there is no duality gap, and the minimum in \eqref{eq:dual} is attained; otherwise $v_r=+\infty$.
\end{enumerate}
\end{theorem}

\begin{proof}
(i) By \textbf{(H)}, $\one\in\intr\sQ_r$, so $\bc+\tau\one=\tau(\one+\bc/\tau)\in\intr\sQ_r$ for $\tau$ large. Then $\bc-A(\tau\bq^\circ)=(\bc+\tau\one)+\tau(-A(\bq^\circ)-\one)\in\intr\sQ_r+\sQ_r\subset\intr\sQ_r$.

(ii) If $\by$ is feasible, $0\le\ell_\by(-A(\bq^\circ)-\one)=-b(\bq^\circ)-\sum_k\ell_{\by_k}(1)$. By Proposition~\ref{prop:H}, each $\by_k$ belongs to $\ell_{\by_k}(1)\,\sB_r(\bg_k)$ with $\sB_r(\bg_k)$ compact, and the feasible set is closed.

{
(iii) For $\be\in V_r$ define the perturbation value function
\[
w(\be):=\sup\{b(\bp):\bp\in P_r,\ \bc+\be-A(\bp)\in\sQ_r\},
\qquad w(\mathbf0)=v_r.
\]
The argument in (i), with $\bc$ replaced by $\bc+\be$, proves feasibility
for every $\be$, so $w(\be)>-\infty$ everywhere.
Convex combinations of feasible points for perturbations $\be_1,\be_2$
are feasible for the corresponding convex combination of perturbations;
taking suprema proves concavity of $w$.
Moreover, if $\bp_0$ is strictly feasible at $\be=\mathbf0$, then it remains
feasible for all sufficiently small $\norm{\be}$, so
$w(\be)\ge b(\bp_0)$ on a neighborhood of $\mathbf0$.
 
Assume first that $v_r<+\infty$ (and note that $v_r>-\infty$ by (i)).
We claim that $w$ is finite everywhere. Indeed, if $w(\be)=+\infty$,
choose feasible points $\bp_i$ for perturbation $\be$ with
$b(\bp_i)\to+\infty$, and fix one feasible point $\bp_-$ for perturbation
$-\be$. Then $(\bp_i+\bp_-)/2$ is feasible at $\mathbf0$, with objective
tending to $+\infty$, a contradiction.
Thus $-w$ is a finite convex function on the whole finite-dimensional
space $V_r$. Its subdifferential at $\mathbf0$ is nonempty
\cite[Theorem~23.4]{roc70}. Equivalently, $w$ has a supergradient there:
there is a tuple $\by$ such that
\[
w(\be)\le v_r+\ell_\by(\be)\qquad\text{for every }\be\in V_r.
\]
 
If $\be\in\sQ_r$, every feasible point at $\mathbf0$ remains feasible
at $\be$, because $\sQ_r+\sQ_r\subseteq\sQ_r$. Hence $w(\be)\ge v_r$,
and the supergradient inequality gives $\ell_\by(\be)\ge0$ on $\sQ_r$.
Therefore $\by\in\sM_r$.
Next, for an arbitrary $\bp\in P_r$, take $\be=A(\bp)-\bc$.
The point $\bp$ is feasible for this perturbation, with zero conic residual,
and consequently
\[
b(\bp)\le w(A(\bp)-\bc)
\le v_r+A^*(\by)(\bp)-c(\by).
\]
Thus the linear functional $b-A^*(\by)$ is bounded above on the entire
vector space $P_r$. It must vanish, since a nonzero linear functional is
unbounded above along a suitable ray. This proves $A^*(\by)=b$.
Taking $\bp=0$ in the last inequality also gives $c(\by)\le v_r$.
Weak duality gives the reverse inequality, so $c(\by)=v_r$:
the dual minimum is attained and there is no gap.
 
Finally, if the dual is feasible, weak duality bounds $v_r$ above by the
finite objective value of any dual feasible point; together with (i),
this makes $v_r$ finite and the preceding argument applies.
If the dual is infeasible, that argument rules out a finite $v_r$;
since (i) rules out $v_r=-\infty$, necessarily $v_r=+\infty$.}

\end{proof}

For polynomial optimization, \textbf{(R)} holds trivially (Section~\ref{sec:pop}) and Theorem~\ref{thm:duality} is the strong duality result of \cite[Theorem~1]{jh16}, whose proof rests on the boundedness of the pseudo-moments under a ball constraint \cite[Lemma~3]{jh16}; here it is extended to box generators, and more generally to \textbf{(H)}. Hypotheses \textbf{(H)} and \textbf{(R)} are the exact-arithmetic counterparts of the two standard ingredients of strong duality proofs in the moment-SOS hierarchy: pseudo-moments bounded by the masses, and bounded masses.

\begin{theorem}[Optimal repair]\label{thm:loss}
Assume \textbf{(H)}, \textbf{(R)}, and that \eqref{eq:dual} is feasible. Then for all $\beps\in\R^m_+$,
\begin{equation}\label{eq:support}
\loss_r(\beps)=\max\Big\{\sum_{k=1}^m\varepsilon_k\,\ell_{\by_k}(1):\ \by\in\sM_r,\ A^*(\by)=b\Big\}\ \le\ -b(\bq^\circ)\,\max_k\varepsilon_k .
\end{equation}
In particular $\loss_r$ is finite, sublinear and nondecreasing. Moreover, if $\hat\bp$, $\beps$ are as in Proposition~\ref{prop:repaired} and $\bq\in P_r$ is rational and feasible for \eqref{eq:rec} with rational SOS multipliers, then $\hat\bp+\bq$ is feasible for \eqref{eq:primal} with rational SOS multipliers, and
\[
b(\hat\bp)+b(\bq)\ \le\ v_r\ \le\ v^\star .
\]
\end{theorem}

\begin{proof}
Hypotheses \textbf{(H)} and \textbf{(R)} do not involve $\bc$, and the dual feasible set does not depend on $\bc$. Apply Theorem~\ref{thm:duality} with $\bc$ replaced by $-\beps$: $-\loss_r(\beps)=\min\{-\sum_k\varepsilon_k\ell_{\by_k}(1)\}$ over the dual feasible set. The inequality in \eqref{eq:support} follows from Theorem~\ref{thm:duality}(ii), or from the feasibility of $\bq=(\max_k\varepsilon_k)\,\bq^\circ$. The last claim was proved after \eqref{eq:rec}.
\end{proof}

The optimal loss is therefore the support function, in the
direction $\beps$, of the set of mass vectors of the dual
feasible pseudo-moments. The maximum in \eqref{eq:support}
is attained by compactness.

A direction can itself be certified by reserving a strictly
positive rational slack vector $\boldsymbol\eta$ and solving
\[
-A(\bq)-\beps-\boldsymbol\eta\in\sQ_r.
\]
Assume that $\beps$ is rational. Rounding and repair applied
to this auxiliary problem produce a rational direction
$\hat\bq$ and rational repair constants
$\boldsymbol\delta\ge0$ such that
\[
-A(\hat\bq)-\beps-\boldsymbol\eta
+\boldsymbol\delta\in\sQ_r,
\]
with rational SOS multipliers. If
$\delta_k\le\eta_k$ for every block, then
\[
-A(\hat\bq)-\beps
=
\bigl(-A(\hat\bq)-\beps-\boldsymbol\eta
      +\boldsymbol\delta\bigr)
+(\boldsymbol\eta-\boldsymbol\delta)
\in\sQ_r.
\]
Thus the reserved slack absorbs the repair, and no further
recession direction is needed. Hypotheses \textbf{(H)} and \textbf{(R)}
ensure strict feasibility of the auxiliary problem;
the inequalities $\delta_k\le\eta_k$ are checked in
exact arithmetic before accepting the certificate.

\section{Examples}\label{sec:examples}

In the examples below, the equality $A^*(\by)=b$ can be solved explicitly for one block of pseudo-moments. The dual \eqref{eq:dual} then becomes a linear matrix inequality problem in free variables, which is the dual form of semidefinite solvers. The multiplier of the matrix inequality returned by the solver is block diagonal and consists of the Gram matrices of \eqref{eq:primal}.
In our examples, we used CSDP, a standard semidefinite solver \cite{bor99}.
The Python scripts are available at 
\url{https://homepages.laas.fr/henrion/software/certquadmod}

\subsection{Polynomial optimization}\label{sec:pop}

Let $f\in\QQ[\bx]$ and let $\bg$ contain ball or box generators. The problem is
\[
v_r=\sup_{v\in\R}\ v\quad\text{s.t.}\quad f-v\in\sQ_r(\bg),
\]
with $m=1$, $P_r=\R$, $b(v)=v$, $A(v)=v$, $c=f$, and dual $\min\{\ell_\by(f):\by\in\sM_r(\bg),\ \ell_\by(1)=1\}$. Here $v_r\le v^\star=\min_{\sK(\bg)}f$. Hypothesis \textbf{(R)} holds with $q^\circ=-1$, and since $\ell_\by(1)=1$ on the dual feasible set, \eqref{eq:support} gives $\loss_r(\varepsilon)=\varepsilon$, attained at $q=-\varepsilon$. The certificate reads
\[
f-(\hat v-\varepsilon)=(\hat e+\varepsilon u_0)+\sum_{j\ge1}g_j\,(\hat s_j+\varepsilon u_j),\qquad \min_{\sK(\bg)}f\ge\hat v-\varepsilon .
\]

\begin{example}
Let $n=1$, $f=x$, $g_1=1-x^2$ and $r=1$, so that $v_1=-1$ with $x+1=\frac12(1+x)^2+\frac12(1-x^2)$; the Gram matrix of $s_0$ is singular. Suppose that rounding produces $\hat v=-1$ and $\hat s_1=\frac{4999}{10000}$. Then $\hat e=\frac{5001}{10000}+x+\frac{4999}{10000}x^2$ has a unique Gram matrix, of determinant $-10^{-8}$: it is not SOS. Theorem~\ref{thm:ball} gives $1=\frac12(1+x^2)+\frac12(1-x^2)$, and with $\varepsilon=2\cdot10^{-8}$ the Gram matrix of $\hat e+\varepsilon u_0$ has positive diagonal and determinant $10^{-16}>0$. Hence $\min_{[-1,1]}x\ge-1-2\cdot10^{-8}$ is certified by
$x+1+\varepsilon=(\hat e+\frac\varepsilon2(1+x^2))+(1-x^2)(\frac{4999}{10000}+\frac\varepsilon2)$.
\end{example}

\subsection{Volume}\label{sec:volume}
Let $K=\sK(\bg_K)\subset B=\sK(\bg_B)$, where $\bg_B$ consists of the generators of a ball or a box $B$ and $\bg_K\supset\bg_B$. Let $\lambda_B$ be the Lebesgue measure on $B$ normalized to a probability measure, and $\blambda_B$ its vector of moments up to degree $2r$, which is rational\footnote{For the unit ball, $\int\bz^{2\ba}\,d\lambda_B(\bz)=\prod_i(2a_i-1)!!\,/\prod_{i=1}^{|\ba|}(n+2i)$.}. The volume problem of \cite{hls09}, written as a supremum, is
\[
v_r=\sup_{v\in\R[\bx]_{2r}}\ -\!\int v\,d\lambda_B\quad\text{s.t.}\quad v-1\in\sQ_r(\bg_K),\quad v\in\sQ_r(\bg_B),
\]
with $b(v)=-\int v\,d\lambda_B$, $A(v)=(-v,-v)$ and $\bc=(-1,0)$. A feasible $v$ dominates the indicator function of $K$ on $B$, hence $\lambda_B(K)\le-v_r$. The dual is
\[
\min\ -\ell_{\by_K}(1)\quad\text{s.t.}\quad \by_K+\by_B=\blambda_B,\quad \by_K\in\sM_r(\bg_K),\ \by_B\in\sM_r(\bg_B).
\]
Hypothesis \textbf{(R)} holds with $q^\circ=1$, for which $-A(q^\circ)-\one=\mathbf 0$ and $-b(q^\circ)=1$.

Since $\ell_{\by_K}(1)+\ell_{\by_B}(1)=1$ on the dual feasible set, where $\ell_{\by_K}(1)$ ranges over $[0,-v_r]$, formula \eqref{eq:support} gives
\[
\loss_r(\beps)=\varepsilon_B+\max(\varepsilon_K-\varepsilon_B,0)(-v_r)\ \le\ \max(\varepsilon_K,\varepsilon_B).
\]
The upper bound is the loss of the constant direction $q=\max(\varepsilon_K,\varepsilon_B)$, whose certificates are trivial; it equals $\loss_r(\beps)$ when $\varepsilon_K\le\varepsilon_B$. The certified statement is
\[
\lambda_B(K)\ \le\ \int\hat v\,d\lambda_B+\max(\varepsilon_K,\varepsilon_B),
\]
with $\hat v+q-1\in\sQ_r(\bg_K)$ and $\hat v+q\in\sQ_r(\bg_B)$ certified by Proposition~\ref{prop:repaired}.

\begin{example}\label{ex:volume}
	Let $n=1$, $B=[-1,1]$ with $\bg_B=(1-x^2)$, and $K=[0,\frac12]=\{g\ge0\}$ with $g=1-(4x-1)^2$, so that $\lambda_B(K)=\frac14$ and $\bg_K=(1-x^2,\,g)$. We solved the moment side with CSDP in dual form, the free variables being the $2r+1$ entries of $\by_K$, with $\by_B=\blambda_B-\by_K$ substituted; the multiplier returned by CSDP yields the five Gram matrices and $v=\sum_jg_{B,j}s_{B,j}$. 
	
{
At order $r=8$ CSDP returned the value $0.3859455$.
Rerunning the accompanying scripts with eight-digit rounding gives
$\norm{D_K}_F\approx3.05\cdot10^{-6}$ and
$\norm{D_B}_F\approx9.39\cdot10^{-7}$.
Lemma~\ref{lem:repair} with the partition of Theorem~\ref{thm:ball}
($\delta_K=\delta_B=\frac19$) yields
\[
\varepsilon_K=\frac{27468}{10^9},\qquad
\varepsilon_B=\frac{8451}{10^9},\qquad
\lambda_B(K)\le\frac{1136786528557}{2945250000000}
\le0.3859728474.
\]
Here $\int\hat v\,d\lambda_B=113670562843/294525000000$.
Both polynomial identities and the positive semidefiniteness of the rational
Gram matrices were checked exactly. The certification loss is negligible
relative to the relaxation error: the bound still has a relative error of
about $54\%$. At order $r=8$, the polynomial $\hat v$ has degree at most
$2r=16$ and approximates the discontinuous indicator function from above.
Comparisons using the same rounded data are reported in
Section~\ref{sec:comparisons}.
}
\end{example}

\subsection{Volume with Stokes constraints}\label{sec:stokes}

The hierarchy of Section~\ref{sec:volume} converges slowly, because $v$ approximates a discontinuous indicator function from above. Following \cite{tlh23}, let $K=\{\bx:g(\bx)\ge0\}\subset B$ for a single $g\in\QQ[\bx]$, where $\{g>0\}$ is nonempty with closure $K$ and $\nabla g\neq0$ on $\partial K=\{g=0\}$. Stokes' theorem provides linear constraints linking $\lambda_K$ to a measure on $\partial K$, whose multiplier is a polynomial vector field $\bw\in\R[\bx]^n$. Let $d_g:=\lceil\deg g/2\rceil$, $\bg_K:=(\bg_B,g)$ and $\bg_{\partial K}:=(\bg_B,g,-g)$. The SOS strengthening of \cite[eq.~(14) and Section~6.1]{tlh23}, written as a supremum, is
\[
v_r=\sup_{(v,\bw)\in P_r}-\!\int v\,d\lambda_B\quad\text{s.t.}\quad
v-\operatorname{div}\bw-1\in\sQ_r(\bg_K),\quad v\in\sQ_r(\bg_B),\quad -\bw\cdot\nabla g\in\sQ_r(\bg_{\partial K}),
\]
with $P_r:=\{(v,\bw):\deg v\le2r,\ \deg\bw\le2r,\ \deg(\bw\cdot\nabla g)\le2r\}$, so that $b(v,\bw)=-\int v\,d\lambda_B$, $A(v,\bw)=(-v+\operatorname{div}\bw,\,-v,\,\bw\cdot\nabla g)$ and $\bc=(-1,0,0)$. Since the multipliers of $g$ and $-g$ are SOS of the same degree, their difference is an arbitrary polynomial, and $\sQ_r(\bg_{\partial K})=\sQ_r(\bg_B)+g\,\R[\bx]_{2(r-d_g)}$. 
Indeed, for $s:=r-d_g$, every polynomial
$p\in\R[\bx]_{2s}$ has a symmetric Gram representation
$p=\bphi_s^TH\bphi_s$.
For sufficiently large $t>0$, both $H+tI$ and $tI$
are positive semidefinite, so
\[
p=\bphi_s^T(H+tI)\bphi_s-\bphi_s^T(tI)\bphi_s
\in\Sigma_s-\Sigma_s.
\]
The three blocks contain $\bg_B$, so \textbf{(H)} holds by Proposition~\ref{prop:H}. If $(v,\bw)$ is feasible, the Gauss formula, with outward normal $-\nabla g/\norm{\nabla g}$ and surface measure $\sigma$ on $\partial K$, gives
\[
\int v\,d\lambda_B\ \ge\ \int_K(1+\operatorname{div}\bw)\,d\lambda_B
=\lambda_B(K)+\frac{1}{\operatorname{vol}B}\int_{\partial K}\frac{-\bw\cdot\nabla g}{\norm{\nabla g}}\,d\sigma\ \ge\ \lambda_B(K),
\]
hence $\lambda_B(K)\le-v_r$, and this bound is at least as good as the one of Section~\ref{sec:volume}, recovered with $\bw=\mathbf 0$. In the variables $\by=(\by_K,\by_B,\by_{\partial K})$ the dual reads
\[
\min\ -\ell_{\by_K}(1)\quad\text{s.t.}\quad\by_K+\by_B=\blambda_B,\qquad
\ell_{\by_K}(\operatorname{div}\bw)+\ell_{\by_{\partial K}}(\bw\cdot\nabla g)=0\quad\forall\,(0,\bw)\in P_r,
\]
with $\by_K\in\sM_r(\bg_K)$, $\by_B\in\sM_r(\bg_B)$ and $\by_{\partial K}\in\sM_r(\bg_{\partial K})$. Here $\by_{\partial K}$ relaxes the boundary measure $\sigma/(\norm{\nabla g}\operatorname{vol}B)$, and the second family of equalities are the Stokes constraints \cite[eq.~(12)]{tlh23}. As before $\by_B$ is explicit, whereas the Stokes constraints are eliminated by linear algebra before calling a solver in dual form.

\paragraph{Finding the recession direction.} With $\beps=(\varepsilon_K,\varepsilon_B,\varepsilon_{\partial K})$, a direction $\bq=(q_v,\bq_\bw)$ is feasible for \eqref{eq:rec} if and only if
\begin{equation}\label{eq:stokesrec}
	q_v-\operatorname{div}\bq_\bw-\varepsilon_K\in\sQ_r(\bg_K),\qquad
	q_v-\varepsilon_B\in\sQ_r(\bg_B),\qquad
	-\bq_\bw\cdot\nabla g-\varepsilon_{\partial K}\in\sQ_r(\bg_{\partial K}),
\end{equation}
and its loss is $\int q_v\,d\lambda_B$. The guide is the dual reading of \textbf{(R)} in Theorem~\ref{thm:duality}(ii): a recession direction is a test function which bounds the masses. Constants bound $\ell_{\by_K}(1)+\ell_{\by_B}(1)=1$, as in Section~\ref{sec:volume}, but they do not see the boundary mass $\ell_{\by_{\partial K}}(1)$, which enters the dual only through the Stokes constraints. It is bounded by testing these constraints with a vector field whose flux through $\partial K$ is positive: by the third condition in \eqref{eq:stokesrec}, $\bq_\bw$ must point strictly outward on $\partial K$. The first condition then prices this field, since $q_v$ must dominate its divergence on $K$. Two outward fields are available.

\emph{Gradient field.} Under the standing assumption $\nabla g\neq0$ on $\partial K$, take $\bq_\bw$ proportional to $-\nabla g$. Let $\gamma,\beta\in\QQ_{>0}$ be such that
\begin{equation}\label{eq:gammabeta}
	\norm{\nabla g}^2-\gamma\in\sQ_r(\bg_{\partial K}),\qquad\beta+\Delta g\in\sQ_r(\bg_K),
\end{equation}
that is, certified bounds $\min_{\partial K}\norm{\nabla g}^2\ge\gamma$ and $\max_K(-\Delta g)\le\beta$. These are two polynomial optimization problems, certified as in Section~\ref{sec:pop} with the trivial recession direction. By Putinar's Positivstellensatz \cite{put93}, every $\gamma<\min_{\partial K}\norm{\nabla g}^2$ and every $\beta>\max_K(-\Delta g)$ are admissible at some finite order, the modules being Archimedean. Then
\[
\bq_\bw:=-\frac{\varepsilon_{\partial K}}{\gamma}\,\nabla g,\qquad q_v:=\max\Big(\varepsilon_B,\ \varepsilon_K+\frac\beta\gamma\,\varepsilon_{\partial K}\Big)
\]
satisfy \eqref{eq:stokesrec}, because $-\bq_\bw\cdot\nabla g-\varepsilon_{\partial K}=\frac{\varepsilon_{\partial K}}{\gamma}(\norm{\nabla g}^2-\gamma)$ and $q_v-\operatorname{div}\bq_\bw-\varepsilon_K=(q_v-\varepsilon_K-\frac\beta\gamma\varepsilon_{\partial K})+\frac{\varepsilon_{\partial K}}{\gamma}(\beta+\Delta g)$. In particular \textbf{(R)} holds with $\bq^\circ=(1+\beta/\gamma,\,-\nabla g/\gamma)$ at every order for which \eqref{eq:gammabeta} holds, and the certified statement is
\[
\lambda_B(K)\ \le\ \int\hat v\,d\lambda_B+\max\Big(\varepsilon_B,\ \varepsilon_K+\frac\beta\gamma\,\varepsilon_{\partial K}\Big).
\]

\emph{Euler field.} Let $g=1-h(\bx-\bx_K)$ with $h$ a form and $\bx_K\in\QQ^n$, as for the Euclidean balls and the $\ell^4$ disk of \cite[Section~6]{tlh23}. Euler's identity gives $(\bx-\bx_K)\cdot\nabla g=(\deg h)(g-1)$, so that
\[
\bq_\bw:=\frac{\varepsilon_{\partial K}}{\deg h}\,(\bx-\bx_K),\qquad q_v:=\max\Big(\varepsilon_B,\ \varepsilon_K+\frac{n}{\deg h}\,\varepsilon_{\partial K}\Big)
\]
satisfy \eqref{eq:stokesrec} with trivial certificates: $-\bq_\bw\cdot\nabla g-\varepsilon_{\partial K}=-\varepsilon_{\partial K}\,g\in\sQ_r(\bg_{\partial K})$, and $\operatorname{div}\bq_\bw=\frac{n}{\deg h}\varepsilon_{\partial K}$ is constant. No auxiliary problem is needed, and the loss is exact:

\begin{proposition}\label{prop:stokesloss}
	If $g=1-h(\bx-\bx_K)$ with $h$ a form, then for all $\beps=(\varepsilon_K,\varepsilon_B,\varepsilon_{\partial K})\in\R^3_+$,
	\[
	\loss_r(\beps)=\varepsilon_B+\max\Big(\varepsilon_K+\frac{n}{\deg h}\,\varepsilon_{\partial K}-\varepsilon_B,0\Big)(-v_r)\ \le\ \max\Big(\varepsilon_B,\ \varepsilon_K+\frac{n}{\deg h}\,\varepsilon_{\partial K}\Big).
	\]
\end{proposition}

\begin{proof}
	Let $\by$ be dual feasible. Since $\pm g\in\sQ_r(\bg_{\partial K})$, we have $\ell_{\by_{\partial K}}(g)=0$. Testing the Stokes constraints with $\bw=\bx-\bx_K$, for which $\operatorname{div}\bw=n$, gives $n\,\ell_{\by_K}(1)+(\deg h)\big(\ell_{\by_{\partial K}}(g)-\ell_{\by_{\partial K}}(1)\big)=0$, hence $\ell_{\by_{\partial K}}(1)=\frac{n}{\deg h}\,\ell_{\by_K}(1)$. Moreover $\ell_{\by_B}(1)=1-\ell_{\by_K}(1)$, and $\ell_{\by_K}(1)$ ranges over $[0,-v_r]$: it vanishes at $\by=(\mathbf 0,\blambda_B,\mathbf 0)$ and its maximum is $-v_r$ by Theorem~\ref{thm:duality}. Formula \eqref{eq:support} gives the equality, and the inequality follows from $-v_r\le1$.
\end{proof}

The identity $\ell_{\by_{\partial K}}(1)=\frac{n}{\deg h}\ell_{\by_K}(1)$ is the pseudo-moment counterpart of the relation between the boundary measure and the volume of a homogeneous star-shaped body; for a Euclidean ball of radius $\rho_K$ it is $\sigma(\partial K)=n\operatorname{vol}(K)/\rho_K$. Compared with Section~\ref{sec:volume}, the acceleration by Stokes constraints costs one more repair, weighted by $n/\deg h$. For a disk, $\norm{\nabla g}^2-\gamma=-\gamma g$ and $\beta+\Delta g=0$ with $\gamma=4/\rho_K^2$ and $\beta=2n/\rho_K^2$, so that the gradient and Euler fields coincide. In both cases the three repairs use the same closed-form partition of unity of $\sQ_r(\bg_B)$, extended by zero to the generators $g$ and $\pm g$.

\begin{example}\label{ex:stokes}
	We continue Example~\ref{ex:volume}. Here $g=1-h(x-x_K)$ with the form $h(u)=16u^2$ and $x_K=\frac14$, so that $n/\deg h=\frac12$, and $\partial K=\{0,\frac12\}$. Since $\partial K$ is finite, the equalities $M_{r-1}(g\,\by_{\partial K})=0$ force $\by_{\partial K}$ to be the moment vector of $a\,\delta_0+c\,\delta_{1/2}$, and the boundary block reduces to $a\ge0$, $c\ge0$. The $2r$ Stokes constraints $\ell_{\by_K}((x^i)')+\ell_{\by_{\partial K}}(x^ig')=0$, $i=0,\dots,2r-1$, are eliminated by an exact rational null space computation, which leaves only three free variables for the semidefinite solver. The multiplier $w$ of the eliminated constraints is recovered from the matrix $X$ returned by the semidefinite solver, by solving in the least squares sense the linear system expressing stationarity with respect to the eliminated variables.
	
	On the SOS side the boundary block is certified by interpolation: with $\hat\pi_0:=\max(-\hat w(0)g'(0),0)$ and $\hat\pi_{1/2}:=\max(-\hat w(\frac12)g'(\frac12),0)$, we let $\hat s_{\partial K,0}:=\hat\pi_0(1-2x)^2+\hat\pi_{1/2}(2x)^2$, $\hat s_{\partial K,1}:=0$, and the multiplier of $g$ is the exact quotient of $-\hat wg'-\hat s_{\partial K,0}$ by $g$. The remainder, of degree at most one, is the mismatch.
{
It vanishes unless $-\hat wg'$ is negative at a boundary point.
Such a sign error may occur near the optimum, where $w$ vanishes on
$\partial K$; in the run reported below the remainder is zero.
 
At order $r=8$ CSDP returned $0.2676380$, with boundary masses
$a=c\approx0.0669095$, consistent with
Proposition~\ref{prop:stokesloss}. Rerunning the scripts with eight-digit
rounding gives $\norm{D_K}_F\approx1.28\cdot10^{-6}$,
$\norm{D_B}_F\approx3.58\cdot10^{-7}$ and
$\norm{D_{\partial K}}_F=0$. Thus
\[
\varepsilon_K=\frac{11529}{10^9},\qquad
\varepsilon_B=\frac{3231}{10^9},\qquad
\varepsilon_{\partial K}=0.
\]
The Euler-field correction has $q_w=0$ and $q_v=\varepsilon_K$, so
\[
\lambda_B(K)\le\frac{3726482074111}{13923000000000}
\le0.2676493626,
\qquad
\int\hat v\,d\lambda_B=\frac{931580388961}{3480750000000}.
\]
All three polynomial identities and the positive semidefiniteness of the
rational Gram matrices were checked exactly. The relative error drops from
about $54\%$ to $7\%$ at the same relaxation order.
}
\end{example}

\subsection{Positivity along D-finite trajectories}\label{sec:dfinite}

Following \cite{ihm26}, let $\bx(t\,|\,\bx_0)\in\R^n$ solve $\dot\bx=F(t)\,\bx+F_0(t)$, $\bx(0)=\bx_0\in\QQ^n$, with $F\in\QQ[t]^{n\times n}$ and $F_0\in\QQ[t]^{n\times1}$, let $f\in\QQ[t,\bx]$, and let $K=[0,1]\times\prod_i[\alpha_i-\rho_i,\alpha_i+\rho_i]$ be a rational box containing the trajectory for $t\in[0,1]$. In the normalized coordinates $z_0=2t-1$, $z_i=(x_i-\alpha_i)/\rho_i$, the generators are $\bg=(1-z_0^2,\dots,1-z_n^2)$. With the Liouville operator $\mathcal Lv:=\partial_tv+(\nabla_\bx v)^T(F(t)\bx+F_0(t))$ and $P_r:=\{v\in\R[t,\bx]_{2r}:\deg\mathcal Lv\le2r\}$, the problem is
\[
v_r=\sup_{v\in P_r}\ v(0,\bx_0)\quad\text{s.t.}\quad f-v\in\sQ_r(\bg),\quad \mathcal Lv\in\sQ_r(\bg),
\]
so that $b$ is a point evaluation, $A(v)=(v,-\mathcal Lv)$ and $\bc=(f,0)$. If $v$ is feasible, then along the trajectory, which stays in $K$, $v\le f$ and $t\mapsto v(t,\bx(t\,|\,\bx_0))$ is nondecreasing, its derivative being $\mathcal Lv\ge0$. Hence
\[
v(0,\bx_0)\ \le\ v^\star=\min_{t\in[0,1]}f(t,\bx(t\,|\,\bx_0)).
\]
In the variables $\by=(\by_{\mathrm{stop}},\by_{\mathrm{occ}})$, the dual reads $\min\,\ell_{\by_{\mathrm{stop}}}(f)$ s.t.\ $\ell_{\by_{\mathrm{stop}}}(v)-\ell_{\by_{\mathrm{occ}}}(\mathcal Lv)=v(0,\bx_0)$ for all $v\in P_r$, an optimal stopping problem: $\by_{\mathrm{stop}}$ relaxes the stopped measure and $\by_{\mathrm{occ}}$ the occupation measure up to the stopping time.

Hypothesis \textbf{(R)} holds with $q^\circ=t-2$, since $\mathcal Lq^\circ-1=0$ and $-q^\circ-1=1-t=(1-t)^2+t(1-t)=\frac14(1-z_0)^2+\frac14(1-z_0^2)\in\sQ_1(\bg)$.

\begin{proposition}
	For all $r\ge1$ and $\beps=(\varepsilon_{\mathrm{stop}},\varepsilon_{\mathrm{occ}})\in\R^2_+$, $\loss_r(\beps)=\varepsilon_{\mathrm{stop}}+\varepsilon_{\mathrm{occ}}$, attained at $q_{\beps}:=\varepsilon_{\mathrm{occ}}(t-1)-\varepsilon_{\mathrm{stop}}$.
\end{proposition}

\begin{proof}
	We have $\mathcal Lq_{\beps}-\varepsilon_{\mathrm{occ}}=0$ and $-q_{\beps}-\varepsilon_{\mathrm{stop}}=\varepsilon_{\mathrm{occ}}(1-t)\in\sQ_1(\bg)$, so $q_{\beps}$ is feasible with $-b(q_{\beps})=\varepsilon_{\mathrm{stop}}+\varepsilon_{\mathrm{occ}}$. Conversely, the moment vectors $\by_{\mathrm{stop}}$ of the Dirac measure at $(1,\bx(1\,|\,\bx_0))$ and $\by_{\mathrm{occ}}$ of the occupation measure of the trajectory on $[0,1]$ are dual feasible with unit masses, and \eqref{eq:support} gives $\loss_r(\beps)\ge\varepsilon_{\mathrm{stop}}+\varepsilon_{\mathrm{occ}}$.
\end{proof}

The certified polynomial is $v:=\hat v+\varepsilon_{\mathrm{occ}}(t-1)-\varepsilon_{\mathrm{stop}}$: the constant $-\varepsilon_{\mathrm{stop}}$ restores $f\ge v$, the term $\varepsilon_{\mathrm{occ}}t$ restores monotonicity along the flow, and $-\varepsilon_{\mathrm{occ}}$ compensates its effect on $f-v$. Explicitly, with a box partition $\bu$,
\begin{align*}
	f-v&=(\hat e_{\mathrm{stop}}+\varepsilon_{\mathrm{stop}}u_0)+\sum_{j\ge1}g_j\,(\hat s_{\mathrm{stop},j}+\varepsilon_{\mathrm{stop}}u_j)+\varepsilon_{\mathrm{occ}}\big((1-t)^2+t(1-t)\big),\\
	\mathcal Lv&=(\hat e_{\mathrm{occ}}+\varepsilon_{\mathrm{occ}}u_0)+\sum_{j\ge1}g_j\,(\hat s_{\mathrm{occ},j}+\varepsilon_{\mathrm{occ}}u_j),
\end{align*}
and therefore $\min_{t\in[0,1]}f(t,\bx(t\,|\,\bx_0))\ge\hat v(0,\bx_0)-\varepsilon_{\mathrm{stop}}-\varepsilon_{\mathrm{occ}}$.

\begin{example}
	Consider the instance of \cite{ihm26}: $n=1$, $\dot x=x+\frac59$, $x_0=0$, $f=2tx-x^2+\frac15$ and $K=[0,1]^2$. Then $x(t\,|\,0)=\frac59(e^t-1)\le t$, so $f(t,x(t))=\frac15+x(2t-x)\ge\frac15$ and $v^\star=\frac15$, although $\min_Kf=-\frac45$. We solved the relaxation of order $r=2$ with CSDP, modelling the moment side directly in dual form: the free variables are the $15$ entries of $\by_{\mathrm{occ}}$, and the constraints are the six matrix inequalities $M_{2-d_j}(g_j\,\by_{\mathrm{stop}})\succeq0$, with $\ell_{\by_{\mathrm{stop}}}(v)=v(0,0)+\ell_{\by_{\mathrm{occ}}}(\mathcal Lv)$ substituted, and $M_{2-d_j}(g_j\,\by_{\mathrm{occ}})\succeq0$, $j=0,1,2$. The multiplier returned by CSDP yields the six Gram matrices and $v=f-\sum_jg_js_{\mathrm{stop},j}$. CSDP returned the primal value $0.1999999984$ and the dual value $0.2000000048$, with relative infeasibilities of the order of $10^{-10}$. Neither is a certified bound, since the constraints hold only approximately, and the dual value exceeds $v^\star\ge v_2$. The computed $\by_{\mathrm{occ}}$ has mass $\ell_{\by_{\mathrm{occ}}}(1)\approx5\cdot10^{-5}$, consistently with an optimal stopping time equal to zero. 
{
Rerunning the scripts with seven-digit rounding gives $\hat v(0,0)=\frac15$,
$\norm{D_{\mathrm{stop}}}_F\approx1.106\cdot10^{-7}$ and
$\norm{D_{\mathrm{occ}}}_F\approx3.050\cdot10^{-7}$.
Theorem~\ref{thm:box}, with $N_2=6$, and Lemma~\ref{lem:repair} give
\[
\varepsilon_{\mathrm{stop}}=\frac{666}{10^9},\qquad
\varepsilon_{\mathrm{occ}}=\frac{1836}{10^9},\qquad
\min_{t\in[0,1]}f(t,x(t\,|\,0))\ge
\frac{99998749}{500000000}=0.199997498.
\]
Both polynomial identities and the positive semidefiniteness of the two
$6\times6$ rational Gram matrices were checked exactly. The certified bound
is within $2.502\cdot10^{-6}$ of $v^\star$.
}
\end{example}

{
\subsection{Comparisons of certification methods}\label{sec:comparisons}
	The certification procedure developed in this paper has three steps: (i)
	round the numerical unknown and SOS factors to rational data, (ii) repair each
	block by adding a constant through a partition of unity, and (iii) absorb these
	constants by a recession correction of the unknown.
	We test two implementations of its repair step: the Frobenius sufficient
	condition of Lemma~\ref{lem:repair}, denoted by ``F'', and the sharper
	projected-Gram test discussed in Remark~\ref{rem:rpl}, denoted by ``P''.
	The projection-and-lifting principle used in P comes from existing work,
	as specified below. The contribution of the present framework is the
	partition construction for truncated quadratic modules and the recession
	correction for general linear conic problems, together with its loss
	characterization in Theorem~\ref{thm:loss}.
	
	We compare F and P with the coefficient interval bound, denoted by ``I'', from \cite{magw15} and the
	coefficient absorption subroutine, denoted by ``A'', from \cite{msput}, on the preceding examples and three
	published POPs. These four procedures receive the same rational rounded
	unknown and SOS multipliers, so their losses measure the accuracy cost of
	certifying the same numerical output. 
	
	\paragraph{Implementations of the proposed framework: F and P.}
	For each block, write
	\[
	\hat s_{k,0}=\bphi_r^T\hat G_{k,0}\bphi_r,\qquad
	\rho_k:=\hat e_k-\hat s_{k,0}=\bphi_r^TD_k\bphi_r,
	\qquad H_k:=\hat G_{k,0}+D_k,
	\]
	where $\hat G_{k,0}\succeq0$ exactly, and $D_k$ is obtained by splitting
	each coefficient of $\rho_k$ equally among its Gram entries.
	Let $N=\dim\R[\bz]_r$. For the univariate examples the solver coordinates
	are $x$; for the other examples each box is mapped affinely to $[-1,1]^n$.
	The partitions used in these experiments have
	$u_{k,0}=\bphi_r^T(I/N)\bphi_r$.
	Consequently, either test below certifies
	$\hat e_k+\varepsilon_k u_{k,0}\in\Sigma_r$; adding the other terms of
	the partition then gives
	$c_k-A_k(\hat\bp)+\varepsilon_k\in\sQ_r(\bg_k)$ by the identity in
	Proposition~\ref{prop:repaired}.
	
	\begin{enumerate}[label=(\roman*),leftmargin=*]
		\item \textbf{F: Frobenius repair from Lemma~\ref{lem:repair}.}
		We use the sufficient condition
		\[
		\frac{\varepsilon_k}{N}\ge\norm{D_k}_F,
		\qquad
		\varepsilon_k=\frac{N\lceil10^9\norm{D_k}_F\rceil}{10^9}.
		\]
		The squared inequality is checked in rational arithmetic. It ensures
		$D_k+\varepsilon_k I/N\succeq0$ independently of the positive
		semidefinite contribution $\hat G_{k,0}$. Thus F is the direct
		implementation of the paper's sufficient repair bound, followed by the
		recession correction described below.
		
		\item \textbf{P: projected-Gram repair within the same framework.}
		Instead of bounding $D_k$ alone, we check
		\[
		H_k+\varepsilon_k I/N\succeq0
		\]
		in exact rational arithmetic. We first test $\varepsilon_k=0$; otherwise
		a numerical eigenvalue proposes a shift on the $10^{-9}$ grid, increased
		until the exact semidefiniteness test succeeds. The numerical eigenvalue
		is only a proposal, never the certificate.
		The equal-splitting projection is the construction in
		\cite[Section~3.1, Lemma~2 and Corollary~3]{nwma25}; the subsequent
		spectral shift and conversion to a constant are the lifting mechanism in
		\cite[Section~3.2, Theorem~4 and Lemma~5]{nwma25}, presented there for
		noncommutative polynomial optimization.
		Here we use an exact rational PSD test and apply the resulting constants
		through our partitions and recession directions. P therefore measures
		the benefit of this existing, sharper repair mechanism within the
		proposed framework. F and P are two repair choices in one certification
		procedure.
	\end{enumerate}
	
	\paragraph{Comparators from existing methods: I and A.}
	These comparators retain the same rounded input and replace the block
	repair step as follows.
	\begin{enumerate}[label=(\roman*),leftmargin=*,start=3]
		\item \textbf{I: coefficient interval bound.}
		We implement the residual bound in
		\cite[Section~2.2, equations~(2.4)--(2.5)]{magw15}.
		After scaling the relevant box to $[0,1]^n$, expand
		$\rho_k=\sum_\alpha c_\alpha\by^\alpha$ exactly and set
		$\varepsilon_k=-\sum_\alpha\min(c_\alpha,0)$.
		Since every monomial belongs to $[0,1]$, this proves
		$\rho_k+\varepsilon_k\ge0$ on the box. For the $K$ and boundary
		blocks in the volume examples, we use the smaller box $[0,\frac12]$;
		for the $B$ block we use $[-1,1]$.
		This is the symbolic-numerical residual-bounding stage of the cited
		method.
        It certifies pointwise inequalities and their resulting objective bounds; it does not establish membership in the original truncated quadratic module.
		
		\item \textbf{A: coefficient absorption combined with our partition.}
		We apply the existing \texttt{absorb} subroutine
		\cite[Algorithm~2, lines~2--7]{msput} to the same $\rho_k$.
		Even monomials change the corresponding diagonal budgets; a term
		$c\bz^{\alpha+\beta}$ with a non-even exponent is written using
		$|c|(\bz^\alpha+\operatorname{sign}(c)\bz^\beta)^2/2$,
		debiting both diagonal budgets by $|c|/2$.
		Among admissible splits we choose one with the most balanced total degrees,
		breaking ties by the basis order.
		If $\delta_k$ is the least nonnegative common initial budget making all
		final budgets nonnegative, then
		$\rho_k+\delta_k\sum_{|\alpha|\le r}\bz^{2\alpha}$ is SOS.
		Our partition converts this monomial reserve into the constant repair
		$\varepsilon_k=N\delta_k$, after which we apply the same recession
		correction as for F and P. Thus A is a hybrid comparator combining the
		published absorption subroutine with our partition and recession steps.
		It does not execute the full \texttt{Putinarsos} algorithm from \cite{msput}; that algorithm's perturbed solving stage is included in the separate experiment below.
	\end{enumerate}
	
	\paragraph{Common objective correction.}
	For each method we use the explicit recession corrections of the preceding
	examples: loss $\varepsilon$ for a POP,
	$\max(\varepsilon_K,\varepsilon_B)$ for volume,
	$\max(\varepsilon_B,\varepsilon_K+\varepsilon_{\partial K}/2)$ with
	Stokes constraints, and
	$\varepsilon_{\mathrm{stop}}+\varepsilon_{\mathrm{occ}}$ for the
	trajectory example. For I, these formulas follow from the pointwise
	inequalities and integration, including the Stokes identity.
	Thus all four columns report certified objective losses, although I
	provides a pointwise certificate and F, P and A provide quadratic-module
	certificates. F, P, I and A require no new semidefinite optimization solve
	after the common numerical solution has been obtained.
	
	\begin{center}
		\begin{minipage}{\linewidth}\centering
			\refstepcounter{table}\label{tab:existing-comparison}
			{\small\textbf{Table~\thetable.} Certification losses on the supplied examples.
				F and P implement the partition framework; I and A are the comparators
				defined above. Smaller is better; entries are rounded for display.}\par\smallskip
			\small\setlength{\tabcolsep}{5pt}
			\begin{tabular}{lrrrrr}
				\hline
				& & \multicolumn{2}{c}{Framework variants} & \multicolumn{2}{c}{Comparators}\\
				Instance & $r$ & F & P & I & A\\
				\hline
				POP & 1 & $1.00\,10^{-8}$ & $1.00\,10^{-9}$ & $7.76\,10^{-9}$ & $7.24\,10^{-9}$\\
				Volume & 8 & $2.75\,10^{-5}$ & $3.33\,10^{-7}$ & $4.50$ & $4.97\,10^{-5}$\\
				Volume with Stokes & 8 & $1.15\,10^{-5}$ & $2.58\,10^{-7}$ & $3.48\,10^{-1}$ & $2.40\,10^{-5}$\\
				D-finite & 2 & $2.50\,10^{-6}$ & $2.00\,10^{-9}$ & $1.01\,10^{-5}$ & $3.44\,10^{-6}$\\
				\hline
			\end{tabular}
		\end{minipage}
	\end{center}
	
	The linear POP row uses an actual solver output, with $\hat v=-1$,
	rather than the illustrative perturbation in Section~\ref{sec:pop}.
	In the volume and Stokes examples, F reduces the loss of I by factors
	of about $1.6\cdot10^5$ and $3.0\cdot10^4$, respectively.
	The large interval losses come from expanding the degree-16 residual of
	the $B$ block under $x=2y-1$; they are sensitive to the basis and domain.
	The volume bound from I can of course be capped by the elementary bound
	$\lambda_B(K)\le1$; the table reports the raw repair loss.
	The smaller $K$ block alone favors I over F in both examples.
	Moreover, I and A both improve on F for the linear POP.
	Thus these data show no uniform dominance of F. The smaller losses of P
	show the conservatism of the Frobenius sufficient condition relative to
	checking the full projected Gram matrix within the same framework.
	
	\paragraph{Published polynomial benchmarks.}
	We add the following three box problems, using generators $1-z_i^2$ after
	normalization. These are common dense relaxations for our comparison;
	we do not reproduce the original papers' sparse relaxations or timings.
	The first POP0 is $f=6-z_1^2-2z_1z_2-2z_2^2$ on $[-1,1]^2$,
	from \cite[Example~26]{msput}, with minimum $1$ and $r=1$.
	The other two come from \cite[Section~2.4]{magw15}. Put $a=3969/625$ and
	\begin{align*}
		\Delta(\bx)={}&x_1x_4(x_2-x_1+x_3-x_4+x_5+x_6)\\
		&+x_2x_5(x_1-x_2+x_3+x_4-x_5+x_6)\\
		&+x_3x_6(x_1+x_2-x_3+x_4+x_5-x_6)\\
		&-x_3(x_2x_4+x_1x_5)-x_6(x_1x_2+x_4x_5).
	\end{align*}
	POP1 minimizes $\partial\Delta/\partial x_4$ with $x_4\in[a,8]$ and
	$x_i\in[4,a]$ for $i\ne4$, using $n=6$ and $r=1$.
	POP2--3 minimizes $4x_1\Delta$ with $x_1,x_2,x_3\in[4,a]$ and
	$x_4=x_5=x_6=a$, using $n=3$ and $r=2$.
	All coefficients and box endpoints are rational.
	
	\begin{center}
		\begin{minipage}{\linewidth}\centering
			\refstepcounter{table}\label{tab:published-comparison}
			{\small\textbf{Table~\thetable.} Common-input comparison on published POPs.
				F and P are the framework variants; I and A are the comparators.
				Each method certifies the lower bound $\hat v$ minus its listed loss.}\par\smallskip
			\small\setlength{\tabcolsep}{5pt}
			\begin{tabular}{lcrrrrr}
				\hline
				& & & \multicolumn{2}{c}{Framework variants} & \multicolumn{2}{c}{Comparators}\\
				Instance & $(n,r)$ & $\hat v$ & F & P & I & A\\
				\hline
				POP0 & $(2,1)$ & $1$ & $3.30\,10^{-8}$ & $6.00\,10^{-9}$ & $6.08\,10^{-8}$ & $2.03\,10^{-8}$\\
				POP1 & $(6,1)$ & $-40.32758016$ & $9.10\,10^{-7}$ & $2.00\,10^{-9}$ & $7.75\,10^{-7}$ & $1.14\,10^{-6}$\\
				POP2--3 & $(3,2)$ & $3645.35514992$ & $1.43\,10^{-5}$ & $0$ & $6.66\,10^{-5}$ & $1.77\,10^{-5}$\\
				\hline
			\end{tabular}
		\end{minipage}
	\end{center}
	
	On these three POPs, the ratios I/F are approximately $1.84$, $0.85$ and
	$4.66$. In particular, the coefficient interval bound is slightly sharper
	than the Frobenius bound on POP1. P certifies POP2--3 without an additional
	shift; this does not assert that $\hat v$ is the exact optimum.

	\paragraph{Separate comparator: a fixed-order implementation of \texttt{Putinarsos}.}
	\texttt{Putinarsos} is the existing perturbation-and-absorption algorithm in
	\cite[Algorithm~6]{msput}. In contrast to A, it reserves a positive
	polynomial perturbation before solving a new semidefinite problem, then
	uses \texttt{absorb} to recover an exact certificate from that new output.
	We implement its perturbed decomposition and exact recovery stages
	\cite[Algorithm~6, lines~4--5 and~10--20]{msput} at the same relaxation
	orders as Table~\ref{tab:published-comparison}.
	For a target lower bound $\ell=\hat v-\eta$, set $h=f-\ell$ and
	$t=\sum_{|\alpha|\le r}\bz^{2\alpha}$.
	We solve for an SOS representation of $h-\tau t$, including the redundant
	generators $1-\bz^{2\alpha}$, $0<|\alpha|\le r$.
	In our implementation, we select a representation by maximizing a common
	positive-semidefiniteness margin for the Gram matrices and nonnegative
	redundant-generator coefficients. Cholesky factors and scalar coefficients
	are rounded to denominator $10^{12}$, followed by exact coefficient
	absorption with initial budget $\tau$ in every monomial.
	Acceptance requires an exact polynomial identity and nonnegative rational
	final budgets.
	
	For a reproducible target policy, we try
	$\eta\in\{\varepsilon_F,10\varepsilon_F,100\varepsilon_F,
	1000\varepsilon_F\}$ in that order, with $\tau=\eta/(4N)$, stopping at
	the first exact certificate. Here $\varepsilon_F$ is the unrounded F entry
	underlying Table~\ref{tab:published-comparison}.
	This target policy is an experimental choice. We keep the relaxation order
	and numerical precision fixed, and replace the order search, perturbation
	halving and precision-increase loops of
	\cite[Algorithm~6, lines~2--3, 6--7 and~21--22]{msput} by the stated
	finite list of targets and reserves. The experiment therefore evaluates
	this fixed-order, double-precision implementation of the core stages of
	\texttt{Putinarsos}; rejection of a target does not establish failure of the full
	algorithm.
	
	\begin{center}
		\begin{minipage}{\linewidth}\centering
			\refstepcounter{table}\label{tab:putinarsos}
			{\small\textbf{Table~\thetable.} Accepted certificates from the fixed-order implementation of the
				perturbed solving and exact recovery stages of \texttt{Putinarsos}.
				The solve count includes rejected targets.}\par\smallskip
			\small\setlength{\tabcolsep}{8pt}
			\begin{tabular}{lrrr}
				\hline
				Instance & \shortstack{Additional\\semidefinite solves} & Accepted $\eta$ & Exact lower bound\\
				\hline
				POP0 & 1 & $3.30\,10^{-8}$ & $0.999999967$\\
				POP1 & 1 & $9.10\,10^{-7}$ & $-40.32758107$\\
				POP2--3 & 3 & $1.428\,10^{-3}$ & $3645.35372192$\\
				\hline
			\end{tabular}
		\end{minipage}
	\end{center}
	
	The first two instances match the target certified by F.
	For POP2--3, the first two targets leave a negative absorption budget;
	the third target succeeds. The CSDP run for POP0 reports reduced
	accuracy, but its recovered certificate passes the exact checks.
	These results illustrate the extra solves and sensitivity to numerical
	precision in this implementation. They do not compare against an optimized
	arbitrary-precision implementation of \texttt{Putinarsos} or establish a general
	accuracy ordering between the algorithms.	

\paragraph{Conclusions.}
On the seven tested instances, projected-Gram repair P yields the smallest
certification loss among the four methods applied to the same rounded
input, whereas F has no uniform advantage over I or A. The particularly
large gains over I in the volume and Stokes examples depend on the chosen
basis and domain. The separate \texttt{Putinarsos} experiment illustrates the
behaviour of our fixed-order, fixed-precision implementation and does not
establish an accuracy or runtime ordering for the full algorithm. These
results support combining the existing projection-and-lifting mechanism
with the proposed quadratic-module partitions and recession corrections
to obtain accurate certificates across several problem classes.

}

\section*{Acknowledgements}

The first author is grateful to Mauricio Velasco for insightful exchanges on rational certification of moment-SOS relaxations. 
This work has been supported by the AI Interdisciplinary Institute ANITI funding through the French ``France 2030'' program under the Grant agreement No.~ANR-23-IACL-0002, the European Union’s HORIZON–MSCA-2023-DN-JD programme under the Horizon Europe (HORIZON) Marie Skłodowska-Curie Actions, grant agreement 101120296 (TENORS). 
The authors acknowledge the use of AI for mathematical writing and numerical experiments.

\appendix

\section{Existence and construction of a recession direction}\label{app:recession}

Hypothesis \textbf{(R)} involves neither $b$ nor $\bc$: it is a property of the map $A$ and of the cone $\sQ_r$ alone. It says that the range of $A$ is onto modulo the cone or, dually, that no nonzero pseudo-moment tuple is annihilated by $A^*$.

\begin{proposition}[Recession directions]\label{prop:R}
	Let $L:=A(P_r)\subset V_r$. The following are equivalent:
	\begin{enumerate}[label=(\alph*),topsep=0pt,itemsep=0pt]
		\item $L\cap\intr\sQ_r\neq\emptyset$;
		\item $L+\sQ_r=V_r$, that is, problem \eqref{eq:primal} is feasible for every $\bc\in V_r$;
		\item $\ker A^*\cap\sM_r=\{\mathbf 0\}$;
		\item for every linear functional $b$ on $P_r$, the feasible set of \eqref{eq:dual} is bounded.
	\end{enumerate}
	Each of them implies \textbf{(R)}, and (b) can be strengthened to $L+\intr\sQ_r=V_r$, that is, strict feasibility of \eqref{eq:primal} for every $\bc\in V_r$. Under \textbf{(H)}, hypothesis \textbf{(R)} is equivalent to (a)--(d), and to
	\begin{enumerate}[label=(\alph*),topsep=0pt,itemsep=0pt,start=5]
		\item there is no $\by\in\sM_r$ with $A^*(\by)=0$ and $\ell_\by(\one)=1$.
	\end{enumerate}
\end{proposition}

\begin{proof}
	(a)$\Rightarrow$(b): let $e:=A(\bq)\in\intr\sQ_r$. Since $-e\in L$, the open set $-e+\intr\sQ_r$ is contained in the convex cone $L+\intr\sQ_r$ and contains $\mathbf 0$. A convex cone containing a neighbourhood of the origin is the whole space, so $L+\intr\sQ_r=V_r$, and a fortiori $L+\sQ_r=V_r$.
	
	(b)$\Rightarrow$(c): the dual cone of $L+\sQ_r$ is $L^\perp\cap\sM_r$, and $L^\perp=\ker A^*$ because $\ell_\by(A(\bq))=A^*(\by)(\bq)$ for all $\bq\in P_r$. The dual cone of $V_r$ is $\{\mathbf 0\}$.
	
	(c)$\Rightarrow$(a): suppose that $L\cap\intr\sQ_r=\emptyset$. The subspace $L$ and the nonempty open convex set $\intr\sQ_r$ can be separated \cite[Theorem~11.3]{roc70}: there are $\by\neq\mathbf 0$ and $\alpha\in\R$ with $\ell_\by(q)\le\alpha\le\ell_\by(e)$ for all $q\in L$ and $e\in\intr\sQ_r$. Since $L$ is a subspace, $\ell_\by$ vanishes on $L$ and $\alpha\ge0$. Hence $\ell_\by\ge0$ on $\intr\sQ_r$, and on $\sQ_r$ because $\sQ_r\subset\operatorname{cl}\intr\sQ_r$ \cite[Theorem~6.3]{roc70}. Thus $\by\in\ker A^*\cap\sM_r$ is nonzero, contradicting (c).
	
	(c)$\Leftrightarrow$(d): the feasible set $F_b:=\{\by\in\sM_r:A^*(\by)=b\}$ of \eqref{eq:dual} is closed and convex. If it is nonempty, its recession cone is $\ker A^*\cap\sM_r$ \cite[Corollary~8.3.3]{roc70}, and $F_b$ is bounded if and only if this cone is $\{\mathbf 0\}$ \cite[Theorem~8.4]{roc70}. Conversely, $F_0=\ker A^*\cap\sM_r$ is a cone, bounded only if it is $\{\mathbf 0\}$.
	
	(a)$\Rightarrow$\textbf{(R)}: since $L=-L$, (a) gives $\bq\in P_r$ with $-A(\bq)\in\intr\sQ_r$, hence $-A(\bq)-\tau\one\in\sQ_r$ for some $\tau>0$, and $\bq^\circ:=\bq/\tau$ satisfies \textbf{(R)}.
	
	\textbf{(R)}$\Rightarrow$(a) under \textbf{(H)}: $-A(\bq^\circ)=\one+\big(-A(\bq^\circ)-\one\big)\in\intr\sQ_r+\sQ_r\subset\intr\sQ_r$.
	
	(c)$\Leftrightarrow$(e) under \textbf{(H)}: if $\by\in\sM_r$ is nonzero, some block $\by_k$ is nonzero, and Proposition~\ref{prop:H}(c) gives $\ell_{\by_k}(1)>0$, hence $\ell_\by(\one)>0$. Therefore $\ker A^*\cap\sM_r$ is nonzero if and only if it contains an element of unit mass.
\end{proof}

\begin{remark}[Massless annihilators]\label{rem:massless}
	Without \textbf{(H)}, hypothesis \textbf{(R)} still implies that every $\by\in\ker A^*\cap\sM_r$ is massless: $0\le\ell_\by(-A(\bq^\circ)-\one)=-\ell_\by(\one)\le0$. Hypotheses \textbf{(H)} and \textbf{(R)} thus split the compactness of the dual feasible set into two statements: \textbf{(H)} says that massless pseudo-moments vanish, \textbf{(R)} that annihilators of $A$ are massless. By Proposition~\ref{prop:R}, when \eqref{eq:dual} is feasible and \textbf{(H)} holds, the compactness in Theorem~\ref{thm:duality}(ii) is equivalent to \textbf{(R)}, not only implied by it, and the strict feasibility in Theorem~\ref{thm:duality}(i) holds for every $\bc$. Assertion (c) is also the pseudo-moment version of a statement about measures, see Proposition~\ref{prop:pointwise}(i), which it implies: the moment vector of a nonzero positive measure on $\sK(\bg_k)$ is a nonzero element of $\sM_r(\bg_k)$.
\end{remark}

\begin{remark}[Roles of $b$ and $\bc$]\label{rem:bc}
	The data $\bc$ play no role in the existence of a recession direction, and $b$ only prices it. If \eqref{eq:dual} is feasible, say at $\by$, then $0\le\ell_\by(-A(\bq^\circ)-\one)=-b(\bq^\circ)-\ell_\by(\one)$, so that $b(\bq^\circ)\le0$ and $-b(\bq^\circ)$ is indeed a loss. If instead $b(\bq^\circ)>0$, then $\bp+\tau\bq^\circ$ is feasible for \eqref{eq:primal} together with $\bp$ for all $\tau\ge0$, and $v_r=+\infty$. When \textbf{(R)} fails under \textbf{(H)}, assertion (e) of Proposition~\ref{prop:R} fails too, which provides $\by\in\sM_r$ with $A^*(\by)=0$ and $\ell_\by(\one)=1$. For every $\bq$ feasible in \eqref{eq:rec} and every $\by'$ feasible in \eqref{eq:dual},
	\[
	-b(\bq)=-\ell_{\by'}(A(\bq))=\ell_{\by'}\big(-A(\bq)-\beps\big)+\ell_{\by'}(\beps)\ \ge\ \sum_{k=1}^m\varepsilon_k\,\ell_{\by'_k}(1),
	\]
	and $\by'+\tau\by$ is feasible in \eqref{eq:dual} for all $\tau\ge0$. Hence $\loss_r(\beps)=+\infty$ as soon as $\varepsilon_k>0$ for some block $k$ with $\by_k\neq\mathbf 0$: on those blocks the repair constants cannot be absorbed, whatever the rounding accuracy.
\end{remark}

\begin{proposition}[Pointwise conditions]\label{prop:pointwise}
	Assume \textbf{(H)}, so that each $\sK(\bg_k)$ is compact, see the proof of Theorem~\ref{thm:ac}.
	\begin{enumerate}[label=(\roman*),topsep=0pt,itemsep=0pt]
		\item If \textbf{(R)} holds, then there is $\bq\in P_r$ with $-A_k(\bq)>0$ on $\sK(\bg_k)$ for $k=1,\dots,m$. This condition holds if and only if no nonzero tuple $(\mu_1,\dots,\mu_m)$ of positive measures, with $\mu_k$ supported on $\sK(\bg_k)$, satisfies $\sum_k\int A_k(\bq)\,d\mu_k=0$ for all $\bq\in P_r$.
		\item Conversely, if there is $\bq\in P$ with $-A_k(\bq)>0$ on $\sK(\bg_k)$ for all $k$, and if every $\sQ(\bg_k)$ is Archimedean, then \textbf{(R)} holds at every sufficiently large order.
	\end{enumerate}
\end{proposition}

\begin{proof}
	(i) If $-A(\bq^\circ)-\one\in\sQ_r$, then $-A_k(\bq^\circ)\ge1$ on $\sK(\bg_k)$ for all $k$. For the equivalence, let $E:=\prod_kC(\sK(\bg_k))$ with the supremum norm, let $E_+$ be the cone of tuples of nonnegative functions, whose interior consists of the tuples of strictly positive functions because the sets are compact, and let $L_E$ be the image of $P_r$ under $\bq\mapsto(A_k(\bq)|_{\sK(\bg_k)})_k$, a finite-dimensional, hence closed, subspace of $E$. If $L_E\cap\intr E_+=\emptyset$, separation of the open convex set $\intr E_+$ from $L_E$ \cite[Theorem~3.4]{rud91} gives a nonzero continuous functional vanishing on $L_E$ and of constant sign on $E_+$, and the Riesz representation theorem turns it into a nonzero tuple of positive measures vanishing on $L_E$. Conversely, if $(\mu_k)$ is such a tuple and $-A_k(\bq)\ge\delta_k>0$ on $\sK(\bg_k)$ for all $k$, then $0=\sum_k\int-A_k(\bq)\,d\mu_k\ge\sum_k\delta_k\,\mu_k(\sK(\bg_k))$ forces all $\mu_k$ to vanish.
	
	(ii) Let $\tau>0$ be smaller than $\min_k\min_{\sK(\bg_k)}(-A_k(\bq))$. Then $-A_k(\bq)-\tau>0$ on $\sK(\bg_k)$ and, by Putinar's Positivstellensatz \cite{put93}, $-A_k(\bq)-\tau\in\sQ_{r'}(\bg_k)$ for some $r'$, which can be taken large enough that $\bq\in P_{r'}$. Then $\bq/\tau$ satisfies \textbf{(R)} at order $r'$, and at every order $r\ge r'$ since $P_{r'}\subset P_r$ and $\sQ_{r'}\subset\sQ_r$.
\end{proof}

\begin{remark}[Uniformity in the order]\label{rem:order}
	Proposition~\ref{prop:pointwise}(ii) provides \textbf{(R)} at some uncontrolled order $r'$, whereas certification needs \textbf{(R)} at the relaxation order $r$ actually solved. Directions with trivial certificates, such as constants, or with certificates of order one, such as $q^\circ=t-2$ in Section~\ref{sec:dfinite} and the Euler field in Section~\ref{sec:stokes}, give \textbf{(R)} at every order; the gradient field of Section~\ref{sec:stokes} relies instead on the auxiliary problems \eqref{eq:gammabeta} at the current order. The mechanism behind a failure of \textbf{(R)} is assertion (c) of Proposition~\ref{prop:R}: a block with $A_k=0$ leaves $\by_k$ free in $\sM_r(\bg_k)$; for an autonomous flow on an infinite horizon, with the time variable removed, the moments of an invariant measure supported in the box annihilate the Liouville operator and have positive mass. In Section~\ref{sec:dfinite}, $\mathcal Lt=1$ excludes invariant measures in the time-augmented space, and $t$ is the corresponding test function.
\end{remark}

\paragraph{Finding a recession direction.}
Proposition~\ref{prop:R} reduces the search to a conic feasibility problem in $\bq$ which involves $A$, $\bg$ and $r$ only. Three levels are available.

\emph{Linear algebra.} Let $\mathscr T\subset\sQ_r$ be a finitely generated subcone of elements with explicit certificates, typically the nonnegative constants, the generators $g_{k,j}$, the products $g\,\bx^\ba$ when $\pm g$ are both generators of a block, and a few squares such as $(1\pm z_i)^2$ for a box. Whether $-\one\in L+\mathscr T$ is a linear programming feasibility problem, solvable in rational arithmetic, and its solutions come with trivial certificates. All directions of Section~\ref{sec:examples} are of this kind.

\emph{Semidefinite programming.} Otherwise, let $\one_k\in V_r$ be the tuple with constant $1$ in block $k$ and $0$ elsewhere, so that $\one=\sum_k\one_k$, and solve \eqref{eq:rec} with $\beps=\one_k$ for $k=1,\dots,m$. Rounding and repair with reserved slack, as described at the end of this section, yield rational directions $\hat\bq^{(k)}\in P_r$ with $-A(\hat\bq^{(k)})-\one_k\in\sQ_r$ and rational SOS multipliers. Then, for every $\beps\in\QQ^m_+$,
\[
\hat\bq_{\beps}:=\sum_{k=1}^m\varepsilon_k\,\hat\bq^{(k)}
\]
is feasible for \eqref{eq:rec}, so that $\loss_r(\beps)\le\sum_k\varepsilon_k\,(-b(\hat\bq^{(k)}))$, with equality when the directions are optimal and $\loss_r$ is additive on $\beps$. These $m$ problems depend neither on $\bc$ nor on the floating-point solution: the directions are computed and certified once for a given model and order, before any instance is solved, and reused for every $\bc$ and every repair $\beps$. The direction $\bq^\circ$ of \textbf{(R)} is the case $\beps=\one$, with loss $-b(\bq^\circ)\max_k\varepsilon_k$ as in Theorem~\ref{thm:loss}. The optimal direction for a given $\beps$ costs one more solve of \eqref{eq:rec}, of the size of \eqref{eq:primal}.

\emph{Deciding \textbf{(R)}.} Under \textbf{(H)}, by Proposition~\ref{prop:R}, exactly one of the two semidefinite feasibility problems
\[
\text{(P)}\quad \bq\in P_r,\ \ -A(\bq)-\one\in\sQ_r,
\qquad\qquad
\text{(D)}\quad \by\in\sM_r,\ \ A^*(\by)=0,\ \ \ell_\by(\one)=1,
\]
has a solution. A floating-point solution of (P) is certified as above. A floating-point solution of (D) is a direction along which the dual feasible set is unbounded, by Remark~\ref{rem:bc}; it explains an apparently unbounded dual and signals that the model, not the rounding, has to be modified, for instance by adding a block or a constraint bounding the masses.

\end{document}